\documentclass[12pt,letterpaper,peerreview]{IEEEtran}

\usepackage{setspace}
\usepackage{amssymb,amsmath}
\usepackage{framed}

\usepackage{graphicx}
\usepackage{url}
\usepackage[left=1.0in,right=1.0in,top=1.25in,bottom=1.0in]{geometry}
\usepackage{subfigure}
\usepackage{tabularx}
\usepackage{enumitem}
\usepackage{rotating}

\usepackage{amsfonts}
\usepackage{algorithm}
\usepackage[noend]{algorithmic}
\algsetup{linenosize=\small}

\usepackage{multirow}
\usepackage{booktabs}
\usepackage{longtable}
\usepackage{color}
\usepackage{cancel}
\usepackage{lscape}

\usepackage[bookmarks=true]{hyperref}

\makeatletter
\newcommand{\manuallabel}[2]{\def\@currentlabel{#2}\label{#1}}
\makeatother

\renewcommand{\algorithmicensure}{\textbf{output:}}

\newcommand{\MAIN}[1]{ \renewcommand{\algorithmicensure}{\textbf{main:}} \ENSURE #1 \renewcommand{\algorithmicensure}{\textbf{output:}}}

 \newtheorem{theorem}{Theorem}

 \newtheorem{observation}[theorem]{Observation}
 
 \newtheorem{definition}[theorem]{Definition}

 \newtheorem{corollary}[theorem]{Corollary}
 \newtheorem{lemma}[theorem]{Lemma}

\title{Haplotype Inference for Pedigrees with Few Recombinations}

\author{
B. Kirkpatrick\thanks{B.K. is at the University of Miami, Coral Gables, Forida, USA, and Intrepid Net Computing, Montana, USA \texttt{bbkirk@intrepidnetcomputing.com}}
}

\begin{document}

\maketitle

\begin{abstract}
  Pedigrees, or family trees, are graphs of family relationships that are used to study inheritance.
  A fundamental problem in computational biology is to find, for a pedigree with $n$ individuals genotyped at every site, a set of Mendelian-consistent haplotypes that have the minimum number of recombinations.
  This is an $\mathsf{NP}$-hard problem and some pedigrees can have thousands of individuals and hundreds of thousands of sites.

  \hspace{1em} 
This paper formulates this problem as a optimization on a graph and introduces a tailored algorithm with a running time of $O(n^{(k+2)}m^{6k})$ for $n$ individuals, $m$ sites, and $k$ recombinations.
  Since there are generally only 1-2 recombinations per chromosome in each meiosis, $k$ is small enough to make this algorithm practically relevant.
\end{abstract}



{\bf Keywords.} Pedigrees, haplotype inference, minimum haplotype recombination configuration (MRHC).

\thispagestyle{empty}


\newpage
\setcounter{page}{1}

\section{Introduction}
\label{sec:introduction}

The study of pedigrees is of fundamental interest to several fields: to computer science due the combinatorics of inheritance~\cite{Steel2006,GeigerEtAl2009}, to epidemiology due to the pedigree's utility in disease-gene finding~\cite{Risch1996,Thornton2007} and recombination rate inference~\cite{Coop2008}, and to statistics due to the connections between pedigrees and graphical models in machine learning~\cite{Lauritzen2003}.
The central calculation on pedigrees is to compute the likelihood, or probability with which the observed data observed are inherited in the given genealogy. 
This likelihood serves as a key ingredient for computing recombination rates, inferring haplotypes, and hypothesis testing of disease-loci positions.
State-of-the-art methods for computing the likelihood, or sampling from it, have exponential running times~\cite{Fishelson2005,Abecasis2002,Sobel1996,Geiger2009,Browning2002}.

The likelihood computation with uniform founder allele frequencies can be reduced to the combinatorial {\sc Minimum Recombination Haplotype Configuration (MRHC)} first introduced by Li and Jiang~\cite{LiJiang2005}.
A solution to MRHC is a set of haplotypes that appear with maximum probability.
The MRHC problem is $\mathsf{NP}$-hard, and as such is unlikely to be solvable in polynomial time.


This paper gives an exponential algorithm for the MRHC problem with running time tailored to the required recombinations $O(n^{(k+2)}m^{6k})$ having exponents that only depend on the minimum number of recombinations $k$ which should be relatively small (i.e. one or two recombinations per chromosome per individual per generation).  This is an improvement on previous formulation that rely on integer programming solvers rather than giving an algorithm which is specific to MRHC~\cite{LiJiang2005}.  We also define the minimum-recombination (MR) graph, connect the MR graph to the inheritance path notation and discuss its properties.

The remainder of this paper is organized as follows.
Section~\ref{sec:pedigreeanalysis} introduces the combinatorial model for the pedigree analysis.
Section~\ref{sec:minimumrecombinationgraph} provides a construction of the MR graph.
Finally, Section~\ref{sec:coloringtheminimumrecombinationgraphbyedgebipartization} gives a solution to the MRHC problem based on a coloring of the minimum recombination graph.
Due to space constraints, several algorithms and proofs have been deferred to the extended version of the paper.

\section{Pedigree Analysis}
\label{sec:pedigreeanalysis}
This section gives the background for inferring haplotype configurations from genotype data of a pedigree. 
We use the Iverson bracket notation, so that $[E]$ equals 1 if the logical expression $E$ is true and 0 otherwise.

A \emph{pedigree} is a directed acyclic graph $P$ whose vertex set $I(P)$ is a set of \emph{individuals}, and whose directed arcs indicate genetic inheritance from parent to child.
A pedigree is \emph{diploid} if each of its individuals has either no or two incoming arcs; for example, human, dog, and cow pedigrees are diploid.
For a diploid pedigree~$P$, every individual without incoming arcs is a \emph{founder} of $P$, and every other individual~$i$ is a \emph{non-founder} for which the vertices adjacent to its two incoming arcs are its \emph{parents} $p_1(i),p_2(i)$, mother and father, respectively.
Let $F(P)$ denote the set of founders of~$P$.

In this paper, every individual has genetic data of importance to the haplotype inference problem.
We abstract this data as follows.
A \emph{site} is an element of an ordered set $\{1,\ldots,m\}$.
For two sites $s,t$ in the interval $[1,m]$, their \emph{distance} is $\mathsf{dist}(s,t) = |s - t|$.
For a pedigree $P$, let $n = |I(P)|$ be the number of its individuals.
A \emph{haplotype} $h$ is a string of length $m$ over $\{0,1\}$ whose elements represent binary \emph{alleles} that appear together on the same chromosome.
We use~$p_1$ and $p_2$ to indicate maternal and paternal chromosomes, respectively, and let $h^{p_1}(i),h^{p_2}(i)$ be binary strings that denote the maternal and paternal haplotypes of individual~$i$.
For a site $s$, the maternal (resp. paternal) haplotype of individual~$i$ at site $s$ is the allele $h^{p_1}(i,s)$ (resp. $h^{p_2}(i,s)$) of the string $h^{p_1}(i)$ (resp. $h^{p_2}(i)$) at position~$s$.
A \emph{haplotype configuration} is a matrix $H$ with $m$ columns and $n$ rows, whose entry $H_{rc}$ at row $r$ and column $c$ is the vector $\binom{h^{p_1}(r,c)}{h^{p_2}(r,c)}$.

Haplotype data is expensive to collect; thus, we observe genotype data and recover the haplotypes by inferring the parental and grand-parental origin of each allele.
The genotype of each individual $i$ at each site $s$ is the conflation $g(i,s)$ of the alleles on the two chromosomes: formally,
\begin{equation}
\label{eqn:consistency}
  g(i,s) = \begin{cases}
             h^{p_1}(i,s), & \mbox{if}~h^{p_1}(i,s) = h^{p_2}(i,s),\\
             2, &\mbox{otherwise} \enspace .
           \end{cases}
\end{equation}
Genotype $g(i,s)$ is \emph{homozygous} if $g(i,s) \in \{0,1\}$ and \emph{heterozygous} otherwise.
Let $G$ be the matrix of genotypes with entry $g(i,s)$ at row $i$ and column $s$.
We have defined the genotypes in the generative direction from the haplotypes.
We are interested in the inverse problem of recovering the haplotypes given the genotypes.
For a matrix $G$ having $\eta$ heterozygous sites across all individuals, there are $2^{\eta-1}$ possible configurations satisfying \emph{genotype consistency} given by~\eqref{eqn:consistency}.

Throughout, we assume that Mendelian inheritance at each site in the pedigree proceeds with recombination and without mutation.
This assumption imposes Mendelian consistency rules on the haplotypes (and genotypes) of the parents and children.
For $\ell\in\{1,2\}$, a haplotype $h^{p_\ell}(i)$ is \emph{Mendelian consistent} if, for every site $s$, the allele $h^{p_\ell}(i,s)$ appears in $p_\ell(i)$'s genome as either the grand-maternal allele $h^{p_1}(p_\ell(i),s)$ or grand-paternal allele $h^{p_2}(p_\ell(i),s)$.
Mendelian consistency is a constraint imposed on our haplotype configuration that is in addition to genotype consistency in~\eqref{eqn:consistency}.
From now on, we will define a haplotype configuration as \emph{consistent} if it is both genotype and Mendelian consistent.

For each non-founder $i\in I(P)\setminus F(P)$ and $\ell\in\{1,2\}$, we indicate the \emph{origin} of each allele of $p_\ell(i)$ by the binary variable $\sigma^{p_\ell}(i,s)$ defined~by
\begin{equation}
\label{eqn:origincondition}
  \sigma^{p_\ell}(i,s) = \begin{cases}
                   p_1, &\mbox{if}~h^{p_\ell}(i,s) = h^{p_1}(p_\ell(i), s),\\
                   p_2, &\mbox{if}~h^{p_\ell}(i,s) = h^{p_2}(p_\ell(i), s) \enspace .
                 \end{cases}
\end{equation}
In words, $\sigma^{p_\ell}(i,s)$ equals $p_1$ if~$h^{p_1}(i,s)$ has grand-maternal origin and equals $p_2$ otherwise.
The set $\sigma(s) = \{(\sigma^{p_1}(i,s),\sigma^{p_2}(i,s))~|~i\in I(P)\}$ is the \emph{inheritance path for site $s$}.
A \emph{recombination} is a change of allele between consecutive sites, that is, if $\sigma^{p_\ell}(i,s) \not= \sigma^{p_\ell}(i,s+1)$ for some $\ell\in\{1,2\}$ and~$s\in\{1,\ldots,m-1\}$.
For a haplotype configuration~$H$, $2^{\zeta}$ inheritance paths satisfy~\eqref{eqn:origincondition}, where $\zeta$ is the number of homozygous sites among all parent individuals of the pedigree.
This means that for a genotype matrix $G$, we have at most $O(2^{\eta - 1}2^{\zeta})$ possible tuples $(H,\sigma)$, and this defines the search space for the MRHC problem where the goal is to choose a tuple $(H, \sigma)$ with a minimum number of recombinations represented in $\sigma$.

For a pedigree $P$ and observed genotype data $G$, the formal problem~is:
\begin{center} 
  \framebox[\textwidth]{
  \begin{tabular}{rl}
    \multicolumn{2}{l}{{\sc Minimum Recombination Haplotypes} (MRHC)}\\
  \textit{Input:}      & A pedigree $P$ with genotype matrix $G$. \\
  \textit{Task:}       & Find $h^{p_\ell}(i,s)$ for $i\in I(P),s\in\{1,\ldots,m\},\ell\in\{1,2\}$ minimizing\\
                       & the number of required recombinations, i.e., compute\\
                       & $\textnormal{argmin}_{(H,\sigma)} \sum_{i\in I(P)\setminus F(P)} \sum^{m-1}_{s \ge 1}\sum_{\ell = 1}^2 \mathbb [\sigma^{p_\ell}(i,s) \ne \sigma^{p_\ell}(i,s+1)]$
  \end{tabular}}

\end{center}


\section{Minimum Recombination Graph}
\label{sec:minimumrecombinationgraph}
We now fix a pedigree $P$ and describe a vertex-colored graph $R(P)$, the minimum recombination graph (MR graph) of $P$, which allows us to reduce the MRHC problem on $P$ to a coloring problem on $R(P)$.
The concept of the MR graph was introduced by Doan and Evans~\cite{DoanEvans2010} to model the phasing of genotype data in $P$.
However, our graph definition differs from theirs, because, as we will argue later, their definition does not model all recombinations of all haplotypes consistent with the genotype data.


\subsection{Definition of the Minimum Recombination Graph}
\label{sec:mrgraphdefinition}
Intuitively, the minimum recombination graph represents the Mendelian consistent haplotypes and the resulting minimum recombinations that are required for inheriting those haplotypes in the pedigree: vertices represent genome intervals, vertex colors represent haplotypes on those intervals, and edges represent the potential for inheritance with~recombination.

Formally, the \emph{minimum recombination graph} of $P$ is a tuple $(R(P), \phi, {\cal S})$, where~$R$ is an undirected multigraph,~$\phi$ is a coloring function on the vertices of $R(P)$, and~$\mathcal S$ is a collection of ``parity constraint sets''.
The vertex set $V(R(P))$ of $R(P)$ consists of one vertex $i_{st}$ for each individual $i\in I(P)$ and each genomic interval $1 \le s < t \le m$, plus one \emph{special} vertex~$b$.
A vertex $i_{st}$ is \emph{regular} if sites $s$ and $t$ are contiguous heterozygous sites in individual $i$, and \emph{supplementary} otherwise.
A vertex $i_{st}$ is \emph{heterozygous} (\emph{homozygous}) if $i$ has heterozygous (homozygous) genotypes at both~$s,t$.

\paragraph{Vertex-coloring}
The coloring function $\phi$ assigns to each regular or supplementary vertex $i_{st}$ a color $\phi(i_{st}) \in \{\mathsf{gray}, \mathsf{blue}, \mathsf{red}, \mathsf{white}\}$.
The color of vertex $i_{st}$ indicates the different ``haplotype fragments'' that are Mendelian consistent at sites $s$ and~$t$ in the genome of individual~$i$.
A \emph{haplotype fragment $f(i_{st})$} of a vertex $i_{st}$ at sites $s$ and $t$ is an (unordered) set of two haplotypes which we will write horizontally with sites $s$ and~$t$ side-by-side and the two haplotypes stacked on top of each other.
Let $\Phi(i_{st})$ be the set of haplotype fragments generated by the color assignment of vertex $i_{st}$.  The colors are defined in Table~\ref{tab:vertexcolorrules} 
The \emph{haplotype pair of individual $i$ at sites $s$ and $t$} is the $\{0,1\}$-valued $2 \times 2$-matrix
$H(i,s,t) = \left(\begin{array}{cc}
              h^{p_1}(i,s) & h^{p_1}(i,t)\\
              h^{p_2}(i,s) & h^{p_2}(i,t)
            \end{array}\right)$.
We denote unordered (set) comparison of the haplotype fragments and haplotype pairs by $H(i,s,t) \doteq f(i_{st})$.
Similarly, for set comparison of sets, we write $\{H(i,s,t)|~\forall H\} \doteq \Phi(i_{st})$ where the first set considers all consistent haplotype configurations $H$.
Then the color and genotype of $i_{st}$ precisely represent its haplotype fragments, as defined in Table~\ref{tab:vertexcolorrules}.
\begin{table}
  \label{tab:vertexcolorrules}
  \centering
  \renewcommand{\arraystretch}{1.2}
\begin{tabular}{cccc}
  \toprule
  ~$g(i,s)$~ & ~$g(i,t)$~ & $\{H(i,s,t)| \forall H \} \doteq \Phi(i_{st})$ & ~$\phi(i_{st})$~ \\
  \midrule
     $2$   &    $2$   & $\{\left({0 \atop 1}{1 \atop 0}\right), \left({0 \atop 1}{0 \atop 1}\right) \}$ & $\mathsf{gray}$\\
     $2$   &    $2$   & $\left({0 \atop 1}{1 \atop 0}\right)$                                         & $\mathsf{red}$\\
     $2$   &    $2$   & $\left({0 \atop 1}{0 \atop 1}\right)$                                         & $\mathsf{blue}$\\
     $0$   &    $0$   & $\left({0 \atop 0}{0 \atop 0}\right)$                                         & $\mathsf{blue}$\\
     $1$   &    $1$   & $\left({1 \atop 1}{1 \atop 1}\right)$                                         & $\mathsf{blue}$\\
     $0$   &    $1$   & $\left({0 \atop 0}{1 \atop 1}\right)$                                         & $\mathsf{red}$\\
     $1$   &    $0$   & $\left({1 \atop 1}{0 \atop 0}\right)$                                         & $\mathsf{red}$\\
     \multicolumn{2}{c}{otherwise} & $\{\left({0 \atop 0} {0 \atop 1}\right), \left({0 \atop 1} {0 \atop 0}\right), \left({1 \atop 1} {0 \atop 1}\right),\left({0 \atop 1} {1 \atop 1}\right)\}$ & $\mathsf{white}$ \\
  \bottomrule
  \end{tabular}
  \smallskip
  \caption{Rules for coloring vertex $i_{st}$ of the minimum recombination graph.
           The $\doteq$ symbol denotes a set comparison operation (i.e.,~an unordered comparison of elements).}
\end{table}
For a heterozygous vertex $i_{st}$, its color $\phi(i_{st})$ indicates the relative paternal origin of the heterozygous alleles at sites $s$ and $t$ and corresponds to a haplotype configuration (red and blue have a one-to-one correspondence with the two possible haplotypes for the sites of $i_{st}$).
But these haplotypes are fragmented, and, hence, may or may not be consistent with a single haplotype configuration.
Note that colors may or may not indicate Mendelian consistent haplotype fragments.

\paragraph{Parity constraint sets}
We now describe the collection $\mathcal S$ of parity constraint sets.
The collection $\mathcal S$ contains one set $S$ for each $\mathsf{gray}$ heterozygous supplementary vertex.
A \emph{parity constraint set} is a tuple $(S,\rho_S)$ with  parity color $\rho_S\in \{\mathsf{red},\mathsf{blue}\}$ for even parity, and a set $S$ consisting of a heterozygous supplementary vertex $i_{st}$ and all regular heterozygous vertices $i_{pq}$ such that $s \le p < q \le t$.
Here, $\rho_S = \mathsf{red}$ (resp. $\rho_S = \mathsf{blue}$) indicates even parity of the $\mathsf{red}$ (resp. $\mathsf{blue}$) vertices.
As sites $s,t$ are heterozygous and $i_{st}$ is supplementary, the set $S$ contains at least two regular vertices $i_{sp}$ and either $i_{pt}$ or $i_{qt}$.

Given~$\mathcal S$, every Mendelian consistent haplotype configuration \emph{induces} a vertex coloring $\phi_{\mathcal S}$ of~$R(P)$, defined by
\begin{equation*}
  \phi_{\mathcal S}(i_{st}) = \begin{cases}
                                \phi(i_{st}), & \mbox{if}~\phi(i_{st}) \not= \mathsf{gray},\\
                                \mathsf{red}, & \mbox{if}~\phi(i_{st}) = \mathsf{gray}~\wedge~\exists~H(i,s,t) \doteq \left({0 \atop 1}{1 \atop 0}\right),\\
                                \mathsf{blue}, & \mbox{otherwise} \enspace .
                              \end{cases}
\end{equation*}
However, we need further constraints to guarantee that the coloring $\phi_{\mathcal S}$ has a corresponding Mendelian consistent haplotype configuration.
Intuitively, these constraints ensure that the collection of overlapping haplotype fragments selected by coloring the gray vertices are consistent with two longer haplotypes.

For coloring $\phi_{\mathcal S}$, the number of $\rho_S$-colored vertices in each parity constraint set $(S, \rho_S) \in\mathcal S$ must be even.  When $\rho_S = \mathsf{red}$ it properly models that the gray vertices $i_{pq}$, $s < p < q < t$ with $\phi(i_{pq}) = \mathsf{gray}$ and $\phi_{\mathcal S}(i_{st}) = \mathsf{red}$ indicate alternating alleles 0-1 along the chromosome.
For now, we focus on the case where $\rho_S = \mathsf{red}$ which is the default color for $\rho_S$.
Informally, we want the red-colored gray vertices in the parity constraint set to indicate alternating 0-1 pattern along the haplotype.
Therefore, the color of the unique supplementary vertex in each set $S$ must agree with the pattern indicated by the regular vertices in $S$.
Later we will see that $\rho_S = \mathsf{blue}$ only for particular cases where the $\mathsf{blue}$ vertices are adjacent to $\mathsf{red}$ vertices on edges without recombination, meaning that these $\mathsf{red}$ vertices indicate alternative allele 0-1 along the chromosome.

We call a parity constraint set $S$ \emph{satisfied by $\phi_{\mathcal S}$} if $S$ contains an even number of vertices $i_{pq}$, $s < p < q < t$ with $\phi(i_{pq}) = \mathsf{gray}$  and color $\phi_{\mathcal S}(i_{st}) = \rho_S $; and we call $\mathcal S$ \emph{satisfiable} if there exists a coloring $\phi_{\mathcal S}$ induced by ${\mathcal S}, \phi, H$ such that each set $S \in\mathcal S$ is satisfied.
By definition, a coloring $\phi_{\mathcal S}$ induced by a Mendelian consistent haplotype configuration satisfies all sets $(S, \phi_{\mathcal S}) \in\mathcal S$.
The converse is also true:
\begin{observation}
\label{thm:redbijectionhaplotypeconfigsparitysatisfyingcolorings}
  Any assignment $\phi_{\mathcal S}$ of colors $\mathsf{red}$ and $\mathsf{blue}$ to vertices $i_{pq}$, $s < p < q < t$ with $\phi(i_{pq}) = \mathsf{gray}$ that satisfies all sets of the form $(S, \phi_{\mathcal S}) \in\mathcal S$ represents a Mendelian consistent haplotype configuration $H$.
\end{observation}
In other words, there is a bijection between haplotype configurations and colorings that satisfy the parity constraint sets.  For $\phi_{\mathcal S}=\mathsf{red}$, the justification follows from the 0-1 alternating alleles of gray vertices in any genotype consistent haplotype.  We will see later that in the instances where we have $\rho_S = \mathsf{blue}$, the bijection will also hold.

\paragraph{Edge creation}
It remains to describe the edge set $E(R(P))$ of $R(P)$, which requires some preparation.
Consider a haplotype configuration $H$ and a minimum recombination inheritance path for those haplotypes.
Let $r$ be a recombination that occurs during the inheritance from an individual $i$ to its child $j$ between contiguous sites $q$ and $q+1$.
Let $\ell \in \{1,2\}$ indicate whether $i = p_{\ell}(j)$ is the maternal or paternal parent of $j$.
Then the recombination $r$ of $i$'s haplotypes is indicated in the inheritance path by $\sigma^{p_\ell}(j,q) \ne \sigma^{p_\ell}(j,q+1)$.
Fixing all recombinations $r'\not=r$ in the inheritance path, $r$ can be shifted to the right or to the left in $j$'s inheritance path to produce a new inheritance path which is also consistent with the haplotype configuration $H$.
The \emph{maximal genomic interval} of $r$ is the unique maximal set $[s,t] = \{s,s+1,\ldots,t-1,t\}$ of sites such that $r$ can placed between any contiguous sites $q,q+1$ in the interval with the resulting inheritance path being consistent with $H$.
Since all genotype data is observed, the maximal genomic interval $[s,t]$ of $r$ always means that both $s,t$ are heterozygous sites in the parent $i$, and therefore $[s,t]$ is determined only by the recombination position $q$ and the pair $\{s,t\}$, independent of $H$.
This interval $[s,t]$ is pertinent to which haplotype fragments are represented in $R(P)$, and it is elucidated by the ``min-recomb property'' defined below.

The set $E(R(P))$ will be the disjoint union of the set $E^+$ of \emph{positive} edges and the set $E^-$ of \emph{negative} edges.
An edge $\{u,v\}\in E(R(P))$ will be called \emph{disagreeing} if either $\{u,v\} \in E^+$ and vertices $u,v$ are colored differently, or if $\{u,v\} \in E^-$ and vertices $u,v$ have the same color.
Our goal is to create edges such that $R(P)$ satisfies the ``min-recomb property''.
\begin{definition}
  Let $P$ be a pedigree with $I(P)$ its set of individuals.
  A graph with vertex set $I(P)$ has the \emph{min-recomb property} if for every individual $j \in I(P)$ with parents $p_1(j),p_2(j)$, and every haplotype configuration~$H$ for the genotype data,
    for $\ell\in\{1,2\}$, a recombination between $i=p_\ell(j)$ and~$j$ in the maximal genomic interval $[s,t]$ is in some minimum recombination inheritance path for $H$ if and only if the recombination is represented in the graph by a disagreeing edge incident to vertex $i_{st} = p_{\ell}(i)_{st}$.
\end{definition}

Let $i_{st}$ be a regular vertex of $R(P)$ with $g(i,s) = g(i,t) = 2$ and let $j\in I(P)\setminus F(P)$ be such that $i = p_\ell(j)$.
Then $\phi(i_{st})\in\{\mathsf{gray},\mathsf{blue},\mathsf{red}\}$, and we create edges incident to $i_{st}$ and $j$ depending on their colors and genotypes, according to Table~\ref{table:cases}.
\begin{table}[htpb]
  \centering
  \begin{tabular}{cccc}
    \toprule
    Case & $\phi(p_{3-\ell}(j))$                       & $\phi(j)$                       & edges to create\\
    \midrule
    1 & $\{\mathsf{gray},\mathsf{blue},\mathsf{red}\}$ & $\{\mathsf{gray},\mathsf{blue},\mathsf{red}\}$ & $\{i_{st},j_{st}\},\{p_{3-\ell}(j)_{st},j_{st}\}\in E^+$\\
    2 & $\mathsf{white}$                               & $\{\mathsf{gray},\mathsf{blue},\mathsf{red}\}$ & $\{i_{st},j_{st}\}\in E^+$\\
    3 & $\{\mathsf{gray},\mathsf{blue},\mathsf{red}\}$ & $\mathsf{white}$                               & $\{i_{st},p_{3-\ell}(j)_{st}\}\in E^-$\\
    4 & $\mathsf{white}$                               & $\mathsf{white}$                               & (see text)\\
    \bottomrule
  \end{tabular}
  \smallskip
  \caption{Rules for creating edges of the minimum recombination graph.}
  \label{table:cases}
\end{table}
Note that $R(P)$ is a multigraph, but there is at most one negative edge $\{i_{st}, p_{3-\ell}(j)\}$ for any tuple~$(j,i_{st} = p_\ell(j),p_{3-\ell}(j))$.

It remains to describe the edges to create in Case 4, when $\phi(p_{3 - \ell}(j)) = \phi(j) = \mathsf{white}$.
This will be done according to the following subcases:
\begin{description}
  \item[4(a)] If $p_{3-\ell}(j)$ and $j$ have a common heterozygous site, that is, if $g(p_{3- \ell}(j),s) = g(j,s) = 2$ or $g(p_{3-\ell}(j),t) = g(j,t) = 2$, then there is a unique site $z\in\{s,t\}$ that is heterozygous in both individuals $j$ and $p_{3-\ell}(j)$.
              Let $q(j)\in\{s,s+1,...,t-1,t\} \setminus \{z\}$ be the heterozygous site in $j$ that is closest to $z$, or $q(j) = +\infty$ if no such site exists.
              Similarly, let $q(p_{3-\ell}(j))\in\{s,s+1\ldots,t\} \setminus \{z\}$ be the heterozygous site in $p_{3-\ell}(j)$ that is closest to $z$, or $q(p_{3-\ell}(j))=+\infty$ if no such site exists.
              If $\min\{q(j),q(p_{3-\ell}(j))\} = +\infty$ then vertex $i_{st}$ remains isolated; otherwise, let
              $z_{\min}  =  \min\{z,q\}, z_{\max} = \max\{z,q\}, \bar{z}  = \{s,t\} \setminus \{z\}$,
              and create edges incident to $i_{st}$ according to Table~\ref{tab:case4aedgecreation}.
              \begin{table}[htpb]
                \centering
                \begin{tabular}{ccc}
                  \toprule
                  $\phi(j_{z_{\min}z_{\max}})$                   & $g(i,\min\{q(j),q(p_{3-\ell}(j))\})$ & edge to create\\
                  \midrule
                  $\{\mathsf{blue},\mathsf{red},\mathsf{gray}\}$ & $= g(p_{3-\ell}(j),\bar{z})$          & $\{i_{st}, j_{z_{\min}z_{\max}}\}\in E^+$\\
                  $\{\mathsf{blue},\mathsf{red},\mathsf{gray}\}$ & $\not= g(p_{3-\ell}(j),\bar{z})$      & $\{i_{st}, j_{z_{\min}z_{\max}}\}\in E^-$\\
                  $\mathsf{white}$                               & $=g(p_{3-\ell}(j),\bar{z})$           & $\{i_{st}, p_{3-\ell}(j)_{z_{\min}z_{\max}}\}\in E^-$\\
                  $\mathsf{white}$                               & $\not= g(p_{3-\ell}(j),\bar{z})$      & $\{i_{st}, p_{3-\ell}(j)_{z_{\min}z_{\max}}\}\in E^+$\\
                  \bottomrule
                \end{tabular}
                \smallskip
                \caption{Case 4(a): rules for creating edges incident to a vertex $i_{st}$ with $\min\{q(j),q(p_{3-\ell}(j))\} < +\infty$.}
                \label{tab:case4aedgecreation}
              \end{table}
  \item[4(b)] If $j$ and $p_{3-\ell}(j)$ do not have a heterozygous site at the same position, then either $g(p_{3-\ell}(j),s) = g(j,t) = 2$ or $g(j,s) = g(p_{3-\ell}(j),t) = 2$.
              Let $z \in \{s,t\}$ be such that $g(p_{3-\ell}(j),z)\not=2$ and let $\bar{z} \in \{s,t\}$ be such that $g(j,\bar{z})\not= 2$.
              If $g(p_{3-\ell}(j),z) = g(j,\bar{z})$, create the edge $\{i_{st},b\}\in E^-$, else create the edge $\{i_{st},b\}\in E^+$.
\end{description}

\paragraph{Graph cleanup}
To complete the construction of $R(P)$, we pass through its list of supplementary vertices to remove some of their edges:
this is necessary as some edges adjacent to a supplementary vertex might over-count the number of recombinations; see the example in Figure~\ref{fig:example}.

Let $\{i_{st}, j_{st}\}$ be an edge adjacent to a supplementary gray vertex $i_{st}$ where $i$ is the parent of $j$.
Let $(S(i_{st}), \rho_{S(i_{st})}) \in \mathcal{S}$ be the set containing $i_{st}$.
If all regular vertices $i_{pq}$ in $S(i_{st})$, for $s \le p < q \le t$, are incident to an edge $\{i_{pq},j_{pq}\}$ then the supplementary edge $\{i_{st}, j_{st}\}$ over-counts.
We remove $\{i_{st}, j_{st}\}$ and replace the set $S(i_{st})$ by a set $S(j_{st})$, which has vertices with the same indices as those in $S(i_{st})$ and where the parity constraint is to have an even number of $\bar{\rho}_{S(i_{st})}$ vertices where $\bar{\rho}_{S(i_{st})} = \mathsf{blue}$ if $\rho_{S(i_{st})} = \mathsf{red}$ and $\bar{\rho}_{S(i_{st})} = \mathsf{red}$ if $\rho_{S(i_{st})} = \mathsf{blue}$.
Notice that $j_{st}$ must also be a supplementary vertex, for the condition to be satisfied.

Note that this edge-removal rule does not apply to edges in Case 4, and does not apply to negative edges, as a negative edge $\{i_{st}, j_{st}\}$  adjacent to a supplementary vertex $i_{st}$ has at least one regular vertex $i_{pq}$, $s \le p < q \le t$ in the parity constraint set $S(i_{st})$ for which there is no edge $\{i_{pq},j_{pq}\}$.

\begin{observation}
\label{thm:bijectionhaplotypeconfigsparitysatisfyingcolorings}
  Any assignment $\phi_{\mathcal S}$ of colors $\mathsf{red}$ and $\mathsf{blue}$ to vertices $i_{st}$ with $\phi(i_{st}) = \mathsf{gray}$ that satisfies all parity constraint sets $(S, \rho_S) \in\mathcal S$ represents a Mendelian consistent haplotype configuration $H$.
\end{observation}


Comparing the MR graph $R(P)$ as defined in this section, with the graph $D(P)$ defined by Doan and Evans~\cite{DoanEvans2010}, we find that $D(P)$ fails to properly model the phasing of genotype data; see Section~\ref{sec:mrgraphcomparison} for details.

\subsection{Algorithms}
\label{sec:mrgraphconstruction}

Our motivation for introducing the $\phi$-colored MR graph and parity constraint sets $\mathcal S$ is to model the existence of Mendelian consistent haplotypes for the genotypes in $P$; we formalize this in Lemma~\ref{thm:haplotypesequalcoloring}.
%
%


\begin{lemma}
\label{thm:haplotypesequalcoloring}
\emph{  Given $(R(P),\phi, \mathcal{S})$, there exists a Mendelian consistent haplotype configuration $H$ for the genotypes if and only if there exists a coloring $\phi_{\mathcal S}$ that satisfies all parity constraint sets in $\mathcal S$.}
\end{lemma}
\begin{IEEEproof}
  Given a haplotype configuration $H$, let $\phi_{\mathcal S}$ be a coloring of regular and supplementary vertices in $I(P)$ defined as follows.
  For any vertex $i_{st}\in I(P)$ with $\phi_{\mathcal S}(i_{st}) \not= \mathsf{gray}$, set $\phi_{\mathcal S}(i_{st}) = \phi(i_{st})$.
  For any vertex $i_{st}\in I(P)$ with $\phi_{\mathcal S}(i_{st}) = \mathsf{gray}$ and $H(i,s,t) \doteq \left({0
\atop 1}{1 \atop 0}\right)$, set $\phi_{\mathcal S}(i_{st}) = \mathsf{red}$.
  For any vertex $i_{st}\in I(P)$ with $\phi_{\mathcal S}(i_{st}) = \mathsf{gray}$ and $H(i,s,t) \doteq \left({0 \atop 1}{0 \atop 1}\right)$, set $\phi_{\mathcal S}(i_{st}) = \mathsf{blue}$.
  Then $\phi_{\mathcal S}$ satisfies the parity constraint sets in $\mathcal S$, since each haplotype in $H$ is a contiguous sequence of alleles.

  Conversely, let $\phi_{\mathcal S}$ satisfy the parity constraint sets in $\mathcal S$.
  We generate the haplotype sequences for all individuals by Algorithm~\ref{alg:mrhaplotypes}, which results in the haplotypes from the colored minimum recombination graph.
  For individual $i$ and site $s$, given its genotype $g(i,s)$ the algorithm arbitrarily selects an $\ell\in\{1,2\}$ and obtain haplotype $h^{p_\ell}(i)$ from the graph.
  Recall that the haplotype fragments are unordered, so the symmetry between the first haplotype fragments is broken by arbitrarily selecting the zero allele of the first locus.
  Since the haplotype fragments of all following vertices overlap with the fragments of the previous vertex, all other symmetries are broken by the original choice.
  Then the algorithm sets $h^{p_{3-\ell}}(i) = g(i,s) - h^{p_\ell}(i)$.
  Let $h_{is}$ be the haplotype allele for $i$ at site $s$.
  For the smallest heterozygous site $s_0$ of $i$, setting $h(i,t) = 0$ allows to arbitrarily select one of the haplotypes of $i$.
  To obtain the rest of the haplotype alleles, the loop iterates along the genome setting the alleles as indicated by the colors.
  All gray vertices are used, and since the parity constraints are satisfied by the supplementary vertices, the alleles set by the regular gray vertices and
the supplementary gray vertices are identical.
\end{IEEEproof}


We defined the minimum recombination graph $(R(P), \phi, \mathcal S)$ in terms of the minimum recombination property, proved that such a graph exists and satisfies the coloring property.


First we give Algorithm~\ref{alg:mrhaplotypes} which allows us to obtain the haplotype allele $h(i,s)$ for each individual~$i$ at each site $s$ from the minimum recombination graph $R(P)$ and the coloring $\phi$ of heterozygous vertices.

\begin{algorithm}
  \caption{\label{alg:mrhaplotypes}MR-HAPLOTYPES($R(P), \phi$)}
  \begin{algorithmic}[1]
 	\REQUIRE minimum recombination graph $R(P)$ and a coloring $\phi$ of heterozygous vertices
	\ENSURE haplotype alleles $h(i,s)$ and $h'(i,s)$ for each individual $i$ at each site $s$.
	\MAIN ~\\

	\FORALL{individual $i$}
		\STATE $s_{\min} \leftarrow$ smallest heterozygous site for $i$;
        \STATE $s_{\max} \leftarrow$ largest heterozygous site of $i$;
		\STATE $h(i,s_0) \leftarrow 0$;
		\WHILE {$s < s_{\max}$}
			\STATE $t \leftarrow \min\{t'~|~t' > s, g(i,s) = g(i,t') = 2\}$
			\IF {$\phi(i_{st}) = \mathsf{red}$}
				\STATE $h(i,t) \leftarrow 1 - h(i,s)$;
			\ENDIF
			\IF {$\phi(i_{st}) = \mathsf{blue}$}
				\STATE $h(i,t) \leftarrow h(i,s)$;
			\ENDIF
			\STATE $s \leftarrow t$;
		\ENDWHILE
		\COMMENT {Obtain $h'$ for $i$ by subtracting $h(i)$ from the allelic genotype.}
                \FORALL{site $s$}
                \STATE $h'(i,s) \leftarrow g(i,s) - h(i,s)$;
                \ENDFOR

	\ENDFOR
  \end{algorithmic}
\end{algorithm}

\begin{algorithm}
  \caption{\label{alg:mrnode}MR-VERTEX($i$, $g(i)$, $s$, $t$) }
  \begin{algorithmic}[1]
 	\REQUIRE individual $i$ with allelic genotypes $g(i)$, and sites $s,t$ \\
	\ENSURE  $\emptyset$, or vertex $i_{st}$ and set $\phi(i_{st})$ \\
	\MAIN ~\\
  \IF {$(g(i,s) = 2 \wedge g(i,t) \not= 2) \vee (g(i,s) \ne 2 \wedge g(i,t) = 2)$}
    \RETURN $\emptyset$
  \ENDIF
  \STATE create a supplementary vertex $i_{st}$
	\IF {$g(i,s) = 2$ and $g(i,t) = 2$}
		\STATE $\phi(i_{st}) \leftarrow \mathsf{gray}$
		\IF {Mendelian consistency requires $\phi(i_{st}) = c$ for some $c\in\{\mathsf{red},\mathsf{blue}\}$}
			\STATE $\phi(i_{st}) \leftarrow c$
		\ENDIF
		\STATE create regular vertices $i_{pq}$ for all $p,q\in \{s + 1,\ldots,t-1\}$\\
		\STATE create parity constraint set $S = \{i_{st}\} \cup \{i_{p,q}~|~s < p,q < t\}$ with parity color $\rho_S = \mathsf{red}$\\
	\ENDIF
	\IF {$g(i,s) \ne 2$ or $g(i,t) \ne 2$}
		\IF{ $g(i,s) = g(i,t)$}
			\STATE $\phi(i_{st}) \leftarrow \mathsf{blue}$
		\ELSE
			\STATE $\phi(i_{st}) \leftarrow \mathsf{red}$
		\ENDIF
	\ENDIF
	\RETURN $i_{st}$ and $\phi(i_{st})$;
  \end{algorithmic}
\end{algorithm}

In the rest of this section we discuss how to construct a minimum recombination graph in polynomial time from the genotype data for all individuals in the pedigree $P$.
We make three claims: (1) that the white vertices are irrelevant, (2) that the algorithms we give construct the minimum recombination graph of $P$, and (3) that the algorithms run in polynomial time.


First, consider the white vertices of $(R(P),\phi, \mathcal S)$.
These are not connected to any other vertex of $R(P)$ and are therefore not involved in any recombinations.
They never change their color and are therefore not involved in specifying the haplotype configuration.
Thus, removing the white vertices from $R(P)$ yields a graph that still satisfies the minimum recombination property and the coloring property.
Our algorithms therefore do not create any white vertices.

Second, we claim that Algorithm~\ref{alg:mrgraph} constructs the minimum recombination graph from the given genotype data for all individuals in the pedigree $P$.
In the first \textbf{for}-loop of Algorithm~\ref{alg:mrgraph}, the regular heterozygous vertices are created, and any vertex created after the first \textbf{for}-loop will be a supplementary.
Considering the color $\phi(i)$ of any heterozygous vertex created.
Both the first \textbf{for}-loop of Algorithm~\ref{alg:mrgraph} and that of Algorithm~\ref{alg:mrnode} contain the same conditions based on Mendelian consistency.
If Mendelian consistency requires vertex $i$ to have a particular color $c\in\{\mathsf{red},\mathsf{blue}\}$, then $\phi(i)$ is set to $c$.
By definition of $(R(P),\phi,\mathcal S)$, any heterozygous vertex is colored a particular color if every Mendelian consistent haplotype configuration has the appropriate corresponding haplotypes.
The analysis of all genotype and haplotype possibilities in the proof of Lemma~\ref{thm:mrgraphproperty1} shows that Mendelian consistency criterion is necessary and sufficient to obtain these colors.
The cases show that when considering this vertex as the parent, there are haplotype configurations for both colors of the vertex, regardless of the genotypes of the children.
However, when this vertex  is the child, there are instances where the vertex has a determined color.
These cases in the tables are marked with bold; the disallowed genotype combinations are indicated with MI and by a slash through the offending color with the only feasible color in bold.
Since the table shows all Mendelian consistent genotype possibilities, it follows that any vertex constrained to be a particular color must be constrained by one of the Mendelian compatibility instances in the table.
Therefore these Mendelian consistency cases are necessary and sufficient for initially coloring the heterozygous vertices.

Note that the parity constraint sets add no further coloring constraints to the heterozygous vertices beyond those given by the Mendelian consistency constraints.
To see this, suppose, for the sake of contradiction, that there is a parity constraint set $S\in\mathcal S$ with exactly one vertex $i_{st}$ of color $\phi(i_{st}) = \mathsf{gray}$.
Then in every haplotype configuration $H$, the color $\phi_{\mathcal S}$ is uniquely determined.
Therefore, of all possible haplotype cases in the proof of Lemma~\ref{thm:mrgraphproperty1}, since the only ones having a determined color for a heterozygous vertex are Mendelian consistency cases, then this single gray vertex color must be determined by Mendelian consistency.

It remains to verify that the edges of $R(P)$ are created according to the rules given above.
For this, observe that Algorithm~\ref{alg:mrtrio} is called only for those vertices $i_{st}$ that are heterozygous and regular.
After these vertices have been created, Algorithm~\ref{alg:mrtrio} implements each of the edge creation cases.

\begin{algorithm}
  \caption{\label{alg:mrtrioa}MR-TRIO-A($g$, $(i = p_\ell(j),p_{3-\ell}(j),j)$, $s$, $t$) }
  \begin{algorithmic}[1]
 	\REQUIRE allelic genotypes $g(i)$ for all individuals $i\in I(P)$, a tuple $(i = p_\ell(j),p_{3-\ell}(j),j)$ of individuals, and sites $s,t$ \\
	\ENSURE  positive and negative edges incident to at least two of $i = p_\ell(j),p_{3-\ell}(j),j$.
	\MAIN ~\\
	\COMMENT{observe that vertex $i_{st}$ exists}
	\IF{vertex $p_{3-\ell}(j)_{st}$ does not exist}
		\STATE MR-VERTEX($g$, $p_{3-\ell}(j)_{st}$, $s$,$t$)
	\ENDIF
	\IF{vertex $j_{st}$ does not exist}
		\STATE MR-VERTEX($g$, $j$, $s$,$t$)
	\ENDIF
	\COMMENT{one of four possible cases can happen}
	\IF{both vertices $j_{st},p_{3-\ell}(j)_{st}$ exist}
	  \STATE create edges $\{i_{st}, j_{st}\}, \{p_{3-\ell}(j)_{st}, j_{st}\} \in E^+$
	\ENDIF
	\IF {vertex $j_{st}$ exists and vertex $p_{3-\ell}(j)_{st}$ does not exist}
		\STATE create edge $\{i_{st}, j_{st}\} \in E^+$
	\ENDIF
	\IF {vertex $p_{3-\ell}(j)_{st}$ exists and vertex $j_{st}$ does not exist}
		\IF {$i < j$}
			\STATE create edge $\{i_{st}, p_{3-\ell}(j)_{st}\} \in E^-$
		\ENDIF
	\ENDIF
  \end{algorithmic}
\end{algorithm}

\begin{algorithm}
  \caption{\label{alg:mrtrio}MR-TRIO($g$, $(i = p_\ell(j),p_{3-\ell}(j),j)$, $s$, $t$) }
  \begin{algorithmic}[1]
 	\REQUIRE allelic genotypes $g(i)$ for all individuals $i\in I(P)$, a tuple $(i = p_\ell(j),p_{3-\ell}(j),j)$ of individuals, and sites $s,t$ \\
	\ENSURE  positive and negative edges incident to at least two of $i = p_\ell(j),p_{3-\ell}(j),j$.
	\MAIN ~\\
        \STATE MR-TRIO-A($g$, $(i = p_\ell(j),p_{3-\ell}(j),j)$, $s$, $t$)
	\IF {none of vertices $p_{3-\ell}(j)_{st},j_{st}$ exists}
		\IF{$g(p_{3-\ell}(j),s) = g(j,s) = 2$ or $g(p_{3-\ell}(j),t) = g(j,t) = 2$}
			\STATE $p \leftarrow i$;
			\IF {$g(p_{3-\ell}(j),t) = g(j,t) = 2$}
				\STATE $p \leftarrow j$;
			\ENDIF

            \STATE $q \to \infty$
            \IF{there exists a site $q \in (s,t)$ with $q \ne p$ and $g(j,q) = 2$}
               \STATE $q \leftarrow \textnormal{argmin}_{q'\in \{s,...,t\}\setminus \{p\} :~g(j,q') = 2}\mathsf{dist}(q',p)$
            \ENDIF
            \IF{there exists a site $q\in (s,t)$ with $q \ne p$ and $g(p_{3-\ell},q) = 2$}
              \STATE $q \leftarrow \textnormal{argmin}_{q'\in \{s,...,t\}\setminus \{p\} :~g(p_{3-\ell}(j),q') = 2}\mathsf{dist}(q',p)$.
            \ENDIF

            \STATE $x \leftarrow p$;
            \STATE $y \leftarrow q$;
			\IF {$x > y$}
				\STATE $x \leftarrow q$;
                \STATE $y \leftarrow p$;
			\ENDIF
			\STATE $\alpha \leftarrow \{s,t\} \setminus \{p\}$;
			\STATE MR-VERTEX($g$, $j$, $x$,$y$)
			\IF {$q$ is finite and $j_{xy}$ exists}
				\IF{$g(i,q) = g(p_{3-\ell}(j),\alpha)$}
					\STATE create edge $\{i_{st}, j_{xy}\} \in E^+$
				\ELSE
					\STATE create edge $\{i_{st}, j_{xy}\} \in E^-$
				\ENDIF
			\ENDIF
			\IF {$q$ is finite and $j_{xy}$ does not exist}
				\STATE MR-VERTEX($g$, $p_{3-\ell}(j)$, $x$,$y$)
				\IF{$g(i,q) = g(p_{3-\ell}(j),\alpha)$}
					\STATE create edge $\{i_{st}, p_{3-\ell}(j)_{xy}\} \in E^-$
				\ELSE
					\STATE create edge $\{i_{st}, p_{3-\ell}(j)_{xy}\} \in E^+$
				\ENDIF
			\ENDIF
		\ENDIF
		\IF{$g(p_{3-\ell}(j),s) = g(j,t) = 2$ or $g(p_{3-\ell}(j),t) = g(j,s) = 2$}
			\STATE let $\alpha \in \{s,t\}$ be such that $g(p_{3-\ell}(j),\alpha)\not=2$;
			\STATE let $\beta \in \{s,t\}$ be such that $g(j,\beta) \ne 2$;
			\IF{$g(p_{3-\ell}(j),\alpha) = g(j,\beta)$}
				\STATE create edge $\{i_{st}, b\} \in E^-$
			\ELSE
				\STATE create edge $\{i_{st}, b\} \in E^+$
			\ENDIF
		\ENDIF
	\ENDIF
  \end{algorithmic}
\end{algorithm}

\begin{algorithm}
  \caption{\label{alg:mrgraph}MR-GRAPH($P$)}
  \begin{algorithmic}[1]
 	\REQUIRE pedigree $P$ with allelic genotypes $g(i)$ for all individuals $i\in I(P)$;
	\ENSURE  the minimum recombination graph $(R, \mathcal S, \phi_{\mathcal S})$ of $P$ \\
	\MAIN ~\\
  \STATE set $\mathcal S = \emptyset$;
	\STATE create special vertex $b$;
	\FORALL{individual $i\in I(P)$}
		\FORALL{pair of contiguous heterozygous sites $s,t$}
			\STATE create a regular vertex $i_{st}$ and set $\phi(i_{st}) = \mathsf{gray}$;
			\IF {Mendelian consistency requires $\phi(i_{st}) c$ for some $c\in\{\mathsf{red},\mathsf{blue}\}$}
				\STATE $\phi(i_{st}) \leftarrow c$;
			\ENDIF
		\ENDFOR
	\ENDFOR
	\FORALL{individual $i\in I(P)$}
		\FORALL{regular vertices $i_{st}$}
			\FORALL{vertices $j$ with $p_\ell(j) = i$}
				\STATE MR-TRIO($g$, $(i_{st},p_{3-\ell}(j),j)$, $s$,$t$)
			\ENDFOR
		\ENDFOR
	\ENDFOR
        \FORALL{gray supplementary vertices $i_{st}$}
           \STATE Let $(S(i_{st}), \rho_{S(i_{st})})$ be the PC set for $i_{st}$.
           \FORALL{edges $\{i_{st},j_{st}\}$ with $i$ the parent of $j$}
            \IF{all regular intervals $i_{pq}$ where $s \le p,q \le t$ have edge $\{i_{pq},j_{pq}\}$}
                \STATE edge $\{i_{st},j_{st}\}$ over counts
                \STATE replace $S(i_{st})$ with $S(j_{st})$ where the vertices of the later set have the same indexes as the former set
                \STATE the new PC set has parity as follows.
                \IF {$\rho_{S(i_{st})} = \mathsf{red}$}
                  \STATE $\rho_{S(j_{st})} = \mathsf{blue}$
                \ELSE
                  \STATE $\rho_{S(j_{st})} = \mathsf{red}$
                \ENDIF
            \ENDIF
           \ENDFOR
        \ENDFOR

  \RETURN $(R,\phi,\mathcal S)$;
  \end{algorithmic}
\end{algorithm}

Third, we claim that Algorithm~\ref{alg:mrgraph} runs in time polynomial in $|P|$.
Its running times is determined by the number of vertices that are processed.
Let $n = |I(P)|$ be the number of individuals in $P$, let $m$ be the number of sites, and $c$ be the maximum number of individuals $j$ for any $i$ with $p_\ell(j) = i$.
Then Algorithm~\ref{alg:mrgraph} runs in time $O(cnm)$, since for each individual $i\in I(P)$ there are at most $m$ vertices for contiguous heterozygous sites. For each of those vertices, Algorithm~\ref{alg:mrtrio} is called at most $c$ times, and performs a constant-time edge-creation operation.

\subsection{Properties of the Minimum Recombination Graph}
\label{sec:mrgraphproperties}
We prove basic properties of the minimum recombination graph $(R(P), \phi, \mathcal S)$.


First, there can be multiple colorings of $\mathsf{gray}$ vertices by $\mathsf{red}$ or $\mathsf{blue}$ that satisfy those parity constraints corresponding to a particular choice of haplotypes for all individuals in $P$; this is formalized in Lemma~\ref{thm:mrgraphproperty1}.
\begin{lemma}
 \label{thm:mrgraphproperty1}
   Given $(R(P),\phi,\mathcal S)$, a coloring $\phi'$ of regular and supplementary vertices of $R(P)$ satisfies all parity constraint set in $\mathcal S$ if
   \begin{equation}
   \label{eqn:coloringproperty}
     \phi'(i_{st}) \in \begin{cases}
                        \{\phi(i_{st})\}, &\mbox{if}~\phi(i_{st}) \not= \mathsf{gray}, \mbox{and regular} \\
                        \{\mathsf{red},\mathsf{blue}\}, & \mbox{if}~\phi(i_{st}) = \mathsf{gray}, \mbox{and regular} \\
                        parity(\rho_s) & \mbox{if supplementary}
                     \end{cases}
   \end{equation}
\end{lemma}
\begin{IEEEproof}
  By definition of $\phi$, for any regular vertex $i_{st}$ with $\phi(i_{st}) = \mathsf{gray}$ there exist two haplotype configurations, one in which $i_{st}$ has the red haplotype fragments, $\left({0 \atop 1}{1 \atop 0}\right)$, and one in which $i_{st}$ has the blue haplotype fragments, $\left({0 \atop 1}{1 \atop 0}\right)$.
  In both cases, there exists a haplotype configuration, one represented by $\mathsf{blue}$ and the other by $\mathsf{red}$.  After coloring all the regular vertices, we can select the color of the supplementary vertices to satisfy parity.  Thus, any coloring $\phi'$ obtained from the haplotype fragments that appear in the haplotype configuration and subject to~\eqref{eqn:coloringproperty} satisfies the parity constraint sets.
\end{IEEEproof}

Second, we show that each edge in the graph is necessary, in that there exists a haplotype configuration with the indicated recombination.
\begin{theorem}
\label{thm:existhcinheritancepath}
  For any edge $e = \{i_{st},j_{pq}\}\in E(R(P))$ there exists a haplotype configuration $H$ having a minimum recombination inheritance path with
the recombination indicated by~$e$.
\end{theorem}
\begin{IEEEproof}
  We will create a coloring $\phi'$ that demonstrates the desired result.
  Without loss of generality, suppose that $s \le p,q \le t$.
  We may further assume that $\phi(i_{st}) = \mathsf{gray}$, since $\phi(i_{st}) = \mathsf{gray}$ if $p \ne s$ or $q \ne t$ and otherwise, at least one of $i_{st},j_{pq}$ must be gray.
  If $\phi(j_{pq}) = \mathsf{gray}$ then set $\phi'(j_{pq}) = \mathsf{blue}$, and set $\phi'(j_{pq}) = \phi(j_{pq})$ otherwise.

  We set $\phi'(i_{st})$ according to $\phi'(j_{pq})$, in order to represent a recombination event by $e$.
  If $\{i_{st},j_{pq}\}\in E^-$, set $\phi'(i_{st}) = \phi(j_{pq})$, and if $\{i_{st},j_{pq}\} \in E^+$, set
  \begin{equation*}
      \phi(i_{st}) = \begin{cases}
                        \mathsf{red},  & \textrm{if } \phi(j_{pq}) = \mathsf{blue},\\
                        \mathsf{blue}, & \textrm{otherwise} \enspace.
                     \end{cases}
  \end{equation*}
  In each of the following cases it is not necessary to assign maternal and paternal origin to the haplotypes that we discuss, because their origins can be found by looking at the genders of the parents and the constraints given by their genotypes.
  We simply verify that in each case, the haplotype fragments indicated by the chosen colors require a recombination in the minimum recombination inheritance.

  In some of the cases below, we will slightly abuse the haplotype fragment notation.
  For some vertices $i_{st}$ we will use $f(i_{st}, q)$ with $s<q<t$ to denote a 3-site haplotype fragment on sites $s,q,t$, rather than the typical 2-site haplotype fragment.
  The third site $q$ is a homozygous site that has been inserted, in order, into the $2 \times 2$ haplotype fragment, so that we can more easily note recombinations between incident vertices whose site indices mismatch.
  We will also use the notation $H(i,s,q,t)$ to refer to the $2 \times 3$ matrix which is the restriction of row $i$ to sites $s,q,t$.

\begin{description}
  \item \textbf{Case} $(s=p, t=q)$:
    This situation corresponds to edge creation cases 1-3.
    \begin{enumerate}
      \item If $g(j,s) = g(j,t) = 2$ then $H(j,s,t) \doteq f(j_{st}) = \binom{0~0}{1~1}$.
            If $\{i_{st},j_{st}\}\in E^+$ then $H(i,s,t) \doteq f(i_{st}) = \binom{0~1}{1~0}$; if $\{i_{st},j_{st}\}\in E^-$ then $h(i,s,t)\in \binom{0~0}{1~1}$.
            In both cases there is a recombination required in the minimum recombination inheritance.
      \item If $g(j,s) \not= 2$ and $g(j,t) \not= 2$ and $\phi(j_{st}) = \mathsf{red}$ then $H(j,s,t) \doteq f(j_{st})= \binom{g(j,s)~g(j,s)}{g(j,s)~g(j,s)}$.
            If $\{i_{st},j_{st}\}\in E^+$ then $h(i,s,t) \doteq \binom{0~1}{1~0}$; if $\{i_{st},j_{st}\}\in E^-$ then $h(i,s,t)\in \binom{0~0}{1~1}$.
            In both cases there is a recombination required in the minimum recombination inheritance.
      \item If $g(j,s) \not= 2$ and $g(j,t) \not= 2$ and $\phi(j_{st}) = \mathsf{blue}$ then $H(j,s,t) \doteq f(j_{st}) = \binom{g(j,s)~g(j,t)}{g(j,s)~g(j,t)}$.
            If $\{i_{st},j_{st}\}\in E^+$ then $H(i,s,t) \doteq f(i_{st}) = \binom{0~1}{1~0}$; if $\{i_{st},j_{st}\}\in E^-$ then $H(i,s,t) \doteq f(i_{st}) = \binom{0~0}{1~1}$.
            In both cases there is a recombination required in the minimum recombination inheritance.
    \end{enumerate}
  \item \textbf{Case} $(s = p, t \not= p)$. (The case $s \not= p, t = q$ is symmetric.)
    This situation corresponds to edge creation case 4, by which $\phi(i_{st}) = \phi(j_{pq}) = \mathsf{gray}$.
    There are two subcases, depending on whether $j$ is a co-parent of $i$ or a child of $i$.
    \begin{enumerate}
      \item If $j$ is a co-parent of $i$, then since $\phi(j_{pq}) = \mathsf{blue}$, the haplotype fragments of $j$ are
            \begin{eqnarray*}
		H(j,s,q,t) \doteq f(j_{st},q) = \left( { 0 ~0 ~g(j,t) \atop  0 ~0 ~g(j,t)} \right)
            \end{eqnarray*}
            If $(i_{st}, j_{sq}) \in E^+$ then the haplotype fragments of $i$ are
            \begin{eqnarray*}
		H(i,s,q,t) \doteq f(i_{st},q) = \left( { 0 ~1-g(j,t) ~1 \atop  1 ~1-g(j,t) ~0} \right)
            \end{eqnarray*}
            and the child $c$ of $i$ and $j$ has haplotype fragments
            \begin{eqnarray*}
		H(c,s,q,t) \doteq f(c_{st},q) = \left( { 0 ~1-g(j,t) ~g(j,t) \atop  1 ~1-g(j,t) ~g(j,t)} \right)
            \end{eqnarray*}
            If $(i_{st}, j_{sq}) \in E^-$ then the haplotype fragments of $i$ are
            \begin{eqnarray*}
		H(i,s,q,t) \doteq f(i_{st},q) = \left( { 0 ~g(j,t) ~0 \atop  1 ~g(j,t) ~1} \right)
            \end{eqnarray*}
            and the child $c$ of $i$ and $j$ has haplotype fragments
            \begin{eqnarray*}
		H(c,s,q,t) \doteq f(c_{st}, q) = \left( { 0 ~g(j,t) ~g(j,t) \atop  1 ~g(j,t) ~g(j,t)} \right)
            \end{eqnarray*}
            There is a recombination required in the minimum recombination inheritance of the haplotypes for both edge types.
      \item If $j$ is a child of $i$ then the haplotype fragments of $j$ are
            \begin{eqnarray*}
		H(j,s,q,t) \doteq f(j_{st},q) = \left( { 0 ~0 ~g(j,t) \atop  1 ~1 ~g(j,t)} \right)
            \end{eqnarray*}
 	    If $\{i_{st}, j_{sq}\} \in E^+$ then the haplotype fragments of $i$ are
    	    \begin{eqnarray*}
		H(i,s,q,t) \doteq f(i_{st},q) = \left( { 0 ~g(j,t) ~1 \atop  1 ~g(j,t) ~0} \right)
    	    \end{eqnarray*}
            If $\{i_{st}, j_{sq}\} \in E^-$ then the haplotype fragments of $i$ are
            \begin{eqnarray*}
		H(i,s,q,t) \doteq f(i_{st},q) = \left( { 0 ~1-g(j,t) ~0 \atop  1 ~1-g(j,t) ~1} \right)
            \end{eqnarray*}
            There is a recombination required in the minimum recombination inheritance for both edge types.
    \end{enumerate}
\end{description}
  The cases above have created fragments of haplotypes for individuals $i$ and $j$.

  It remains to prove that there is a coloring of the remaining $\mathsf{gray}$ vertices that satisfies the parity constraint sets.
  We have set the color of at most one vertex in any parity constraint set, since vertices $i_{st}$ and $j_{st}$ belong to different individuals.
  We next argue that any parity constraint set $S$ with a $\mathsf{gray}$ vertex contains at least two $\mathsf{gray}$ vertices; this implies that setting the color of one $\mathsf{gray}$ vertex in $S$ means there exists a coloring of the remaining $\mathsf{gray}$ vertices that satisfies $S$.
  For the sake of contradiction, suppose there is a parity constraint set $S\in\mathcal S$ with a single $\mathsf{gray}$ vertex $x$.
  Then all other vertices in $S$ are heterozygous and colored either $\mathsf{red}$ or $\mathsf{blue}$.
  By definition, those vertices have been colored due to every haplotype configuration containing the haplotype fragments of that color.
  Then the definition of parity constraint tells us that $x$ must be set to a particular color in every haplotype configuration.
  Therefore, $\phi(x) = c$ for some $c\in\{\mathsf{red},\mathsf{blue}\}$, since $\phi(x) = c$ in every Mendelian consistent haplotype configuration.
  This yields the desired contradiction.
\end{IEEEproof}

Third, we prove that $(R,\phi,\mathcal S)$ satisfies the min-recomb~property.
\begin{theorem}
\label{thm:minrecomb}
  Let $H$ be a Mendelian consistent haplotype configuration, let $i,j\in I(P)$ be such that $i = p_\ell(j)$, and let $s,t$ be sites such that $s < t$.
  Then a recombination between $i$ and $j$ in the maximal genomic interval $[s,t]$ is in some minimum recombination inheritance path of $H$ if and only if it is represented in $R(P)$ by a disagreeing edge incident to $i_{st}$.
\end{theorem}
\begin{IEEEproof}
  First, let the recombination of interest be indicated by $s_{c,q}^{p_\ell(i)} \ne s_{c,q+1}^{p_\ell(i)}$ for some $\ell\in\{1,2\}$.
  Then the maximal genomic interval $[s,t]$ for the recombination where $s \le q < t$ is given by the minimum $s_0,t_0$ to $q$ such that $g(i,s_0) = g(i,t_0) = 2$.
  Therefore, to prove the forward direction, we need to ascertain that there is a disagreeing edge adjacent to $i_{st}$.
  We prove this by considering all possible genotypes of $(i = p_\ell(j),p_{3-\ell}(j),j)$ and all possible haplotypes given those genotypes.
  The haplotypes induce colors for the three vertices, and the number of disagreeing edges incident to them must match the minimum number of recombinations for the haplotypes.
  As $i_{st}$ is heterozygous, its genotype is determined.
  However, the colors of $p_{3-\ell}(j)_{st}$ and $j_{st}$ are under-determined and could be one of $\mathsf{gray}, \mathsf{red}, \mathsf{blue}, \mathsf{white}$.
  Depending on whether $p_{3-\ell}(j)_{st}$ and $j_{st}$ are $\mathsf{white}$, we derive the four edge creation cases.
  For each of those four cases, we consider all possible non-white colors for $v_{ij}$ and $c_{ij}$ and then all haplotype possibilities, as is done in Table~\ref{tab:mrgraphproperty1}.
  In every case, the number of disagreeing edges is identical to the number of minimum number of recombinations.
  Edge creation case (4) is the most complicated, because it must consider some three-site haplotypes.
  However, together the subcases of case (4) account for all possible genotypes and Table~\ref{tab:mrgraphproperty1}  considers all possible haplotypes for those genotypes.
  Since every haplotype possibility is represented below, every possible recombination is also represented.
  As in the proof of Lemma~\ref{thm:mrgraphproperty1}, each time a haplotype configuration has a minimum recombination of $i$'s haplotypes (the left vertex in the figure), there is a disagreeing edge adjacent to $i$.
  Thus, in Table~\ref{tab:mrgraphproperty1}, each recombination is represented in $R(P)$ by a disagreeing edge incident to $i_{st}$.

  \medskip

  Second, the converse direction follows from the case analysis in the proof of Lemma~\ref{thm:mrgraphproperty1}.
  Let $\{i_{st},j_{pq}\}$ be a disagreeing edge between some heterozygous vertices $i_{st},j_{pq}$; we prove that there is a recombination of $i$'s haplotypes in the interval $[s,t]$ that was inherited by some child of $i$.
  We may assume that $s = p$; then either $t = q$ or $t > q$.
  \begin{enumerate}
    \item $t = q$ and $j$ is a child of $i$:
      Then in every haplotype configuration where $e$ is disagreeing there is a recombination required between $i$'s haplotypes during inheritance to $j$.
      We conclude that $e$ represents a minimum recombination in the maximal genomic interval $[s,t]$.
    \item $t = q$ and $j,i$ have a common child $c$:
      Let $c_{st}$ be such that $\phi(c_{st}) = \mathsf{white}$.
      This corresponds to edge creation case (3) and that in all cases that there is a disagreeing edge, there is a minimum recombination which can be placed between $i$ and $c$.
    \item $t > q$ and $j$ is a child of $i$:
      Then every haplotype configuration disagreeing on $e$ has a recombination between individual $i$ and $j$.
    \item $t > q$ and $i,j$ have a common child $c$:
      Let $c_{pq}$ be such that $\phi(c_{pq}) = \mathsf{white}$.
      Then in each instance where $H$ induces a disagreement on $e$, there is a minimum recombination which can be placed between $i$ and $c$.
    \item $t > q$ and $j = b$:
      Let $c$ be a child of $i$ such that edge creation case (4.b) is satisfied, that is, $p_{3-\ell}(c) = j$ and $c,j$ do not both have a heterozygous site at the same position at either site $s$ or $t$.
      Then in all instances where the haplotype configuration induces a disagreement on this edge, there is a minimum recombination which must be explained between $i$ and $c$.
  \end{enumerate}

 \medskip

  We finish the proof of Thm~\ref{thm:minrecomb} by haplotype case analysis.
  For sites $s$ and $t$, we look at a triple of vertices mother, father, child and consider their possible genotypes and haplotypes.
  These haplotype cases can be grouped into categories that correspond to edge creation cases of the minimum recombination graph, and they are grouped that way below.
  Within each case, we break the haplotype cases into two varieties: those that are unconstrained by Mendelian consistency and those that are not.
  The varieties constrained by Mendelian consistency have their last column in bold, indicating the only allowed coloring of a particular vertex.
  Haplotypes are written horizontally with columns being SNPs.
  Vertices are arranged in a triangle with two parents above the child, and white vertices are not drawn.
  Positive edges only connect parents and children, and negative edges connect either two parents or a parent and a child.
  Vertex types correspond to haplotypes as follows:

\renewcommand\arraystretch{1.2}

\begin{table}[h!]
  \centering
  \begin{tabular}{ccc}
    \toprule
    genotypes of $i$ at $s,t$           & $\quad\phi(i_{st})\quad$  & haplotype fragments\\
    \midrule
    $g(i,s) = g(i,t) = 2$               & $\mathsf{red}$  & $\left({0 \atop 1}{1 \atop 0}\right)$\\
    $g(i,s) = g(i,t) = 2$               & $\mathsf{blue}$ & $\left({0 \atop 1}{0 \atop 1}\right)$\\
    $2 \not= g(i,s) = g(i,t)$           & $\mathsf{blue}$ & $\left({0 \atop 0}{0 \atop 0}\right)$ or $\left({1 \atop 1}{1 \atop 1}\right)$\\
    $2 \not= g(i,s) \ne g(i,t) \not= 2$ & $\mathsf{red}$  & $\left({0 \atop 0}{1 \atop 1}\right)$ or $\left({1 \atop 1}{0 \atop 0}\right)$\\
    \bottomrule
  \end{tabular}
  \smallskip
  \caption{Correspondence between vertex types and haplotypes}
\end{table}

\noindent

\def\@currentlabel{\roman{5}}\label{tab:mrgraphproperty1}
In Supplement Table~\ref{tab:mrgraphproperty1}, each line corresponds to a vertex coloring of the graph on the left.
Only cases with some Mendelian consistent possibilities are shown.
The colors are for the vertices in the order: left parent, right parent, child.
The codes in the rightmost column indicate the number of minimum recombinations: [0R] means non-recombinant haplotypes, [1R] means one recombinant haplotype, [2R] means two recombinant haplotypes, and [MI] means Mendelian inconsistency.


\end{IEEEproof}

Theorem~\ref{thm:minrecomb} proves that the edge construction cases result in an MR graph, since those particular edges satisfy the min-recomb~property.
\begin{corollary}
  For a Mendelian consistent haplotype configuration $H$, let $\phi'$ be the coloring induced on $R(P)$ by $H$, and let
  $E' = \{\{i_{st}, j_{pq}\} \in E^-~|~\phi'(i_{st}) = \phi'(j_{pq})\} \cup \{\{i_{st}, j_{pq}\} \in E^+~|~\phi'(i_{st}) \ne \phi'(j_{pq})\}$.
  Then the minimum number of recombinations required for any inheritance of those haplotypes equals $|E'|$.
\end{corollary}

Note that similar to the proof of Theorem~\ref{thm:minrecomb}, from $R(P)$ and $\phi$, we can exploit the edge cases for the disagreeing edges to obtain a minimum recombination inheritance path from $R(P)$ in time $O(|E(R(P))|)$ time.
The running time is due to a constant number of cases being considered for each disagreeing edge.
From each of the cases, a feasible inheritance path is an immediate consequence.

\begin{corollary}
 A solution to the {\sc MRHC} problem corresponds to a coloring $\phi_{\mathcal S}$ that satisfies $\mathcal S$ and has a minimum number of disagreeing edges.
\end{corollary}

\subsection{Comparison of the MR Graph with the Doan-Evans Graph}
\label{sec:mrgraphcomparison}
We now compare the MR graph $R(P)$, as defined in Section~\ref{sec:minimumrecombinationgraph}, with the graph $D(P)$ defined by Doan and Evans~\cite{DoanEvans2010}.
We claim that the graph $D(P)$ fails to properly model the phasing of genotype data.


First, in $D(P)$ any vertex that represents two heterozygous sites is colored gray.
However, as some of the gray vertices are constrained by Mendelian consistency to be either red or blue, $D$ represents Mendelian inconsistent haplotype configurations.
For example, in some instances where both parents are white, i.e. $\left({0 \atop 0}{0 \atop 1}\right)$ and $\left({0 \atop 1}{0 \atop 0}\right)$, the heterozygous child must be colored red.

Second, $D(P)$ violates the minimum recombination property: in Figure 1(c) of their paper~\cite{DoanEvans2010}, there exists haplotypes for the two parents and child such that $H$ indicates a different number of recombinations than required by the haplotypes.
Specifically, let the left parent have haplotypes $0101$ and $1110$, the right parent have haplotypes $0010$ and $1111$, and the child have haplotypes $0111$ and $1111$.
Then $D(P)$ indicates one recombination, whereas the minimum number of recombinations required by the haplotypes is two.

\begin{figure}[ht!]
  \centering
    \includegraphics[width=\textwidth]{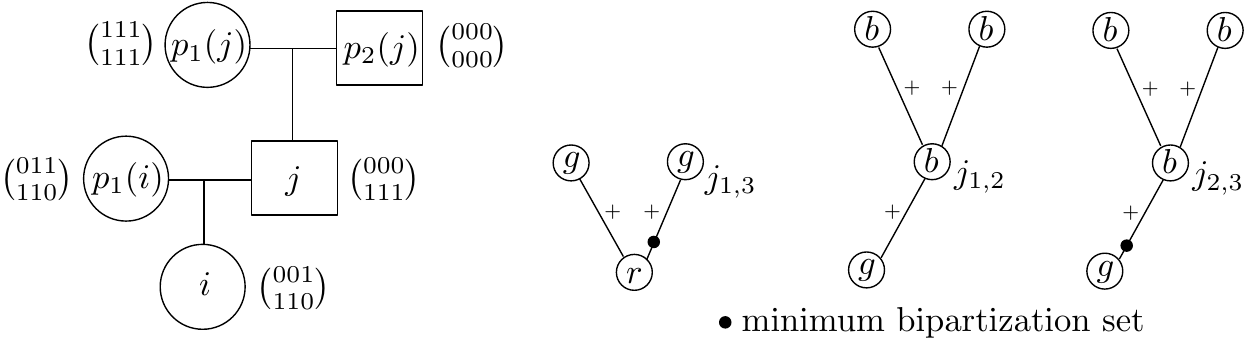}
  \caption{The specified haplotypes induce two disagreeing edges in $D(P)$, but only one recombination is required to inherit the haplotypes.}
  \label{fig:example}
\end{figure}

Third, the parity constraint sets defined by Doan and Evans~\cite{DoanEvans2010} can over-count the number of recombinations.
For example, consider the pedigree $P$ with $n = 5$ individuals consisting of an individual $i$, its parents, and its paternal grand-parents, see Figure~\ref{fig:example}.

\section{Coloring the MR Graph by Edge Bipartization}
\label{sec:coloringtheminimumrecombinationgraphbyedgebipartization}
In this section, we solve a variant of an edge bipartization problem on a perturbation of the minimum recombination graph.
The solution to this problem is in one-to-one correspondence with a Mendelian consistent haplotype configuration for the genotype data, because of Observation~\ref{thm:bijectionhaplotypeconfigsparitysatisfyingcolorings}.

First, we perturb the graph $(R(P), \phi, \mathcal S)$ by substituting each of the positive edges in $R(P)$ by two negative edges.
That is, bisect every positive edge $\{i_{st},j_{st}\}\in E^+$ with a new gray vertex $x$ and add the resulting two edges $\{i_{st},x\},\{x,j_{st}\}$.
Once this step has been completed for all positive edges of $R(P)$, call the resulting graph $R(P)^-$.
Observe that $R(P)^-$ is not a minimum recombination graph, since the new $\mathsf{gray}$ vertices do not represent a maximal genomic interval.
Further, colorings of $R(P)$ and $R(P)^-$ are in one-to-one correspondence, as the color of $i_{st}$ in $R(P)$ equals the color of $i_{st}$ in $R(P)^-$.
Similarly, $R(P)^-$ has the same number of disagreeing edges of a given coloring of $R(P)$, and thus preserves the number of recombinations of any coloring.
Thus, by Observation~\ref{thm:bijectionhaplotypeconfigsparitysatisfyingcolorings}, $R(P)$ has a bipartization set of size $k$ if and only if $R(P)^-$ has.

Second, we perturb the graph $(R(P)^-,\phi,\mathcal S)$ by turning $R(P)^-$ into an uncolored graph $\overline{R(P)}$.
The graph $\overline{R(P)}$ has the same vertex set as $R(P)^-$ (with colors on the vertices removed), plus two additional vertices $v_r$ and $v_b$.
The graph contains all edges of $R(P)^-$, plus a \emph{parity edge}
 for every vertex colored $\mathsf{red}$ connecting it to $v_b$ and a parity edge for every vertex colored $\mathsf{blue}$ connecting it to $v_r$.
This way, color constraints are preserved.
For a graph, a subset $B$ of its edges is called a \emph{bipartization set} if removing the edges in $B$ from the graph yields a bipartite graph.

A bipartization set is \emph{minimal} if it does not include a bipartization set as proper subset. A bipartization set is \emph{respectful} if it also satisfies the parity constraint sets. 
We claim that respectful
bipartization sets of $R(P)^-$ are respectful bipartization set of $\overline{R(P)}$.
Those bipartization sets of $\overline{R(P)}$ that are not bipartization sets of $R(P)^-$ contain at least one parity edge.
Here we need to compute a bipartization set $B$ (with size at most $k$) of non-parity edges such that the graph $R(P) - B$ satisfies all parity constraint sets in $\mathcal S$; we call such a set~$B$ \emph{respectful (with respect to $\mathcal S$)}.

\subsection{The Exponential Algorithm}
A MRHC problem instance has parameters $n$ for the number of individuals, $m$ for the number of sites, and $k$ for the number of recombinations.

The algorithm considers in brute-force fashion the minimum number of recombinations $\{0,1,2,...,k\}$ and stops on the first $k$ such that there exists some set $S$ of $k$ edges whose removal from the graph produces (1) a bipartite graph and (2) satisfies the parity constraints.  For each selection of $k$ edges, the two checks require (1) traversing the graph in a depth-first search in time $O(n^2m^4)$ and (2) computing the parity of all the parity constraint sets in time $O(nm^3)$.

The number of sets $S$ with $k$ recombination edges is $|E|^k$ where $E = E(\overline{R(P)})$ is the edge set of $\overline{R(P)}$ and where $|E| = O(nm^2)$.  So, the running time of the whole algorithm is $O(n^{(k+2)}m^{6k})$.

\section{Discussion}
\label{sec:discussion}
We give an exponential algorithm to compute minimum recombination haplotype configurations for pedigrees with all genotyped individuals, with only polynomial dependence on the number $m$ of sites (which can be very large in practice) and small exponential dependence on the minimum number $k$ of recombinations.
This algorithm significantly improves, and corrects, earlier results by Doan and Evans~\cite{DoanEvans2010,DoanEvans2011}. 
An open question is how this algorithm performs when implemented and applied to data.  Another open question is how to accomodate missing alleles in data.

\section*{Acknowledgment.} BK thanks M. Mnich at the Cluster of Excellence, Saarland University, Saarbr\"{u}cken, Germany for critical reading of the manuscript.


{
\bibliographystyle{splncs03}
\bibliography{pedigree}
}

\begin{IEEEbiographynophoto}{B. Kirkpatrick} received a PhD degree from the University of California Berkeley and was an assistant professor in the Department of Computer Science at the University of Miami during part of this work.  Kirkpatrick is now the owner of an R\&D start-up Intrepid Net Computing.
\end{IEEEbiographynophoto}


\newpage

\section*{Supplement}

\setcounter{page}{1}

\setcounter{table}{4}

\noindent
In Table~\ref{tab:mrgraphproperty1}, each line corresponds to a vertex coloring of the graph on the left.
Only cases with some Mendelian consistent possibilities are shown.
The colors are for the vertices in the order: left parent, right parent, child.
The codes in the rightmost column indicate the number of minimum recombinations: [0R] means non-recombinant haplotypes, [1R] means one recombinant haplotype, [2R] means two recombinant haplotypes, and [MI] means Mendelian inconsistency.

\renewcommand{\arraystretch}{1.0}

\footnotesize
\setlength{\LTcapwidth}{0.9\textwidth}
\begin{center}
  \begin{longtable}{lll|l|l|l}
    \caption{Edge creation rules for the minimum recombination graph}\\
    \toprule
    Case & \begin{sideways}\parbox{15mm}{Colors}\end{sideways} &  Parent & Parent & Child & Rec. \\
    \midrule
    \endfirsthead
    \toprule
     Case & \begin{sideways}\parbox{15mm}{Colors}\end{sideways} &  Parent & Parent & Child & Rec. \\
    \midrule
    \endhead
    \bottomrule
    \endfoot
    \bottomrule
    \endlastfoot

     \multirow{2}{*}{\includegraphics[scale=0.7]{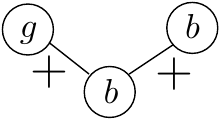}} &  B & $\left({0 \atop 1}{0 \atop 1}\right)$ & $\left({0 \atop 0}{0 \atop 0}\right)$ or $\left({1 \atop 1}{1 \atop 1}\right)$ & $\left({0 \atop 0}{0 \atop 0}\right)$ or $\left({1 \atop 1}{1 \atop 1}\right)$ & 0R  \\

	 & R & $\left({0 \atop 1}{1 \atop 0}\right)$ & $\left({0 \atop 0}{0 \atop 0}\right)$ or $\left({1 \atop 1}{1 \atop 1}\right)$ & $\left({0 \atop 0}{0 \atop 0}\right)$ or $\left({1 \atop 1}{1 \atop 1}\right)$ & 1R \\
\midrule

    	  \multirow{2}{*}{\includegraphics[scale=0.7]{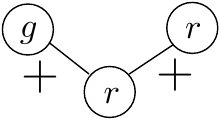}} & B & $\left({0 \atop 1}{0 \atop 1}\right)$ & $\left({0 \atop 0}{1 \atop 1}\right)$ or $\left({1 \atop 1}{0 \atop 0}\right)$ & $\left({0 \atop 0}{1 \atop 1}\right)$ or $\left({1 \atop 1}{0 \atop 0}\right)$ & 1R  \\
       & R & $\left({0 \atop 1}{1 \atop 0}\right)$ & $\left({0 \atop 0}{1 \atop 1}\right)$ or $\left({1 \atop 1}{0 \atop 0}\right)$ & $\left({0 \atop 0}{1 \atop 1}\right)$ or $\left({1 \atop 1}{0 \atop 0}\right)$ & 0R  \\
\midrule
    	  \multirow{2}{*}{\includegraphics[scale=0.7]{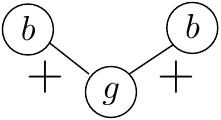}} & \bf{B} &  $\left({0 \atop 0}{0 \atop 0}\right)$ or $\left({1 \atop 1}{1 \atop 1}\right)$ & $\left({0 \atop 0}{0 \atop 0}\right)$ or $\left({1 \atop 1}{1 \atop 1}\right)$ & $\left({0 \atop 1}{0 \atop 1}\right)$ & {\bf 0R }  \\
	 &  \cancel{R} & & & & {\bf MI } \\

      \midrule
    	 \multirow{2}{*}{\includegraphics[scale=0.7]{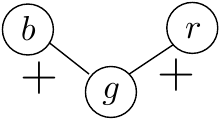}} & B & $\left({0 \atop 0}{0 \atop 0}\right)$ or $\left({1 \atop 1}{1 \atop 1}\right)$ & $\left({0 \atop 0}{1 \atop 1}\right)$ or $\left({1 \atop 1}{0 \atop 0}\right)$ & $\left({0 \atop 1}{0 \atop 1}\right)$ & 1R     \\
	 &  R & $\left({0 \atop 0}{0 \atop 0}\right)$ or $\left({1 \atop 1}{1 \atop 1}\right)$ & $\left({0 \atop 0}{1 \atop 1}\right)$ or $\left({1 \atop 1}{0 \atop 0}\right)$ & $\left({0 \atop 1}{1 \atop 0}\right)$ & 1R    \\
      \midrule
    	 \multirow{2}{*}{\includegraphics[scale=0.7]{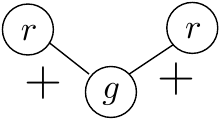}} & \bf{R} & $\left({0 \atop 0}{1 \atop 1}\right)$ or $\left({1 \atop 1}{0 \atop 0}\right)$ & $\left({0 \atop 0}{1 \atop 1}\right)$ or $\left({1 \atop 1}{0 \atop 0}\right)$ & $\left({0 \atop 1}{1 \atop 0}\right)$ & {\bf 0R }   \\
	 &  \cancel{B} & & & & {\bf MI } \\
\midrule
    	\multirow{3}{*}{\includegraphics[scale=0.7]{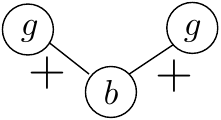}} & BB & $\left({0 \atop 1}{0 \atop 1}\right)$ & $\left({0 \atop 1}{0 \atop 1}\right)$ & $\left({0 \atop 0}{0 \atop 0}\right)$ or $\left({1 \atop 1}{1 \atop 1}\right)$ & 0R\\
	& BR & $\left({0 \atop 1}{0 \atop 1}\right)$ & $\left({0 \atop 1}{1 \atop 0}\right)$ & $\left({0 \atop 0}{0 \atop 0}\right)$ or $\left({1 \atop 1}{1 \atop 1}\right)$ & 1R\\
	& RR & $\left({0 \atop 1}{1 \atop 0}\right)$ & $\left({0 \atop 1}{1 \atop 0}\right)$ & $\left({0 \atop 0}{0 \atop 0}\right)$ or $\left({1 \atop 1}{1 \atop 1}\right)$ & 2R\\

    	 \multirow{3}{*}{ \includegraphics[scale=0.7]{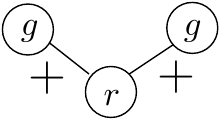}} & BB & $\left({0 \atop 1}{0 \atop 1}\right)$ & $\left({0 \atop 1}{0 \atop 1}\right)$ & $\left({0 \atop 0}{1 \atop 1}\right)$ or $\left({1 \atop 1}{0 \atop 0}\right)$ & 2R\\
	&  BR & $\left({0 \atop 1}{0 \atop 1}\right)$ & $\left({0 \atop 1}{1 \atop 0}\right)$ & $\left({0 \atop 0}{1 \atop 1}\right)$ or $\left({1 \atop 1}{0 \atop 0}\right)$ & 1R\\
	&  RR & $\left({0 \atop 1}{1 \atop 0}\right)$ & $\left({0 \atop 1}{1 \atop 0}\right)$ & $\left({0 \atop 0}{1 \atop 1}\right)$ or $\left({1 \atop 1}{0 \atop 0}\right)$ & 0R\\
\midrule
    	 \multirow{4}{*}{ \includegraphics[scale=0.7]{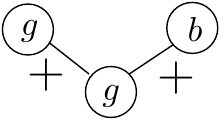}} & B\bf{B} & $\left({0 \atop 1}{0 \atop 1}\right)$ & $\left({0 \atop 0}{0 \atop 0}\right)$ or $\left({1 \atop 1}{1 \atop 1}\right)$ & $\left({0 \atop 1}{0 \atop 1}\right)$ & {\bf 0R } \\
	&  \cancel{BR} & $\left({0 \atop 1}{0 \atop 1}\right)$ & $\left({0 \atop 0}{0 \atop 0}\right)$ or $\left({1 \atop 1}{1 \atop 1}\right)$ & $\left({0 \atop 1}{1 \atop 0}\right)$ & {\bf MI } \\
	& R\bf{B} & $\left({0 \atop 1}{1 \atop 0}\right)$ & $\left({0 \atop 0}{0 \atop 0}\right)$ or $\left({1 \atop 1}{1 \atop 1}\right)$ & $\left({0 \atop 1}{0 \atop 1}\right)$ & {\bf 1R } \\
	 & \cancel{RR} & $\left({0 \atop 1}{1 \atop 0}\right)$ & $\left({0 \atop 0}{0 \atop 0}\right)$ or $\left({1 \atop 1}{1 \atop 1}\right)$ & $\left({0 \atop 1}{1 \atop 0}\right)$ & {\bf MI } \\


    	\newpage

    	  \multirow{4}{*}{\includegraphics[scale=0.7]{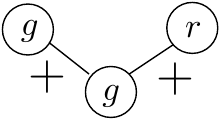}} & \cancel{BB} & $\left({0 \atop 1}{0 \atop 1}\right)$ & $\left({0 \atop 0}{1 \atop 1}\right)$ or $\left({1 \atop 1}{0 \atop 0}\right)$ & $\left({0 \atop 1}{0 \atop 1}\right)$ & {\bf MI}\\
	&  B\bf{R} & $\left({0 \atop 1}{0 \atop 1}\right)$ & $\left({0 \atop 0}{1 \atop 1}\right)$ or $\left({1 \atop 1}{0 \atop 0}\right)$ & $\left({0 \atop 1}{1 \atop 0}\right)$ & {\bf 1R } \\
	& \cancel{RB} & $\left({0 \atop 1}{1 \atop 0}\right)$ & $\left({0 \atop 0}{1 \atop 1}\right)$ or $\left({1 \atop 1}{0 \atop 0}\right)$ & $\left({0 \atop 1}{0 \atop 1}\right)$ & {\bf MI } \\
	 & R\bf{R} & $\left({0 \atop 1}{1 \atop 0}\right)$ & $\left({0 \atop 0}{1 \atop 1}\right)$ or $\left({1 \atop 1}{0 \atop 0}\right)$ & $\left({0 \atop 1}{1 \atop 0}\right)$ & {\bf 0R } \\
\midrule
    	
   \multirow{6}{*}{\includegraphics[scale=0.7]{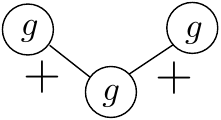}} & BBB & $\left({0 \atop 1}{0 \atop 1}\right)$ & $\left({0 \atop 1}{0 \atop 1}\right)$ & $\left({0 \atop 1}{0 \atop 1}\right)$ & 0R \\
	&  BBR & $\left({0 \atop 1}{0 \atop 1}\right)$ & $\left({0 \atop 1}{0 \atop 1}\right)$ & $\left({0 \atop 1}{1 \atop 0}\right)$ & 2R  \\
	&  BRB & $\left({0 \atop 1}{0 \atop 1}\right)$ & $\left({0 \atop 1}{1 \atop 0}\right)$ & $\left({0 \atop 1}{0 \atop 1}\right)$ & 1R \\
	&  BRR & $\left({0 \atop 1}{0 \atop 1}\right)$ & $\left({0 \atop 1}{1 \atop 0}\right)$ & $\left({0 \atop 1}{1 \atop 0}\right)$ & 1R \\
	&  RRB & $\left({0 \atop 1}{1 \atop 0}\right)$ & $\left({0 \atop 1}{1 \atop 0}\right)$ & $\left({0 \atop 1}{0 \atop 1}\right)$ & 2R \\
	&  RRR & $\left({0 \atop 1}{1 \atop 0}\right)$ & $\left({0 \atop 1}{1 \atop 0}\right)$ & $\left({0 \atop 1}{1 \atop 0}\right)$ & 0R \\

     \midrule
    \includegraphics[scale=0.7]{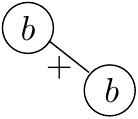}
	&  & $\left({0 \atop 0} {0 \atop 0}\right)$ or $\left({1 \atop 1} {1 \atop 1}\right)$ &
		$\left({0 \atop 0} {0 \atop 1}\right)$, $\left({0 \atop 1} {0 \atop 0}\right)$, $\left({1 \atop 1} {0 \atop 1}\right)$, or $\left({0 \atop 1} {1 \atop 1}\right)$  &
		$\left({0 \atop 0} {0 \atop 0}\right)$ or $\left({1 \atop 1} {1 \atop 1}\right)$ & 0R \\

      \midrule


    \includegraphics[scale=0.7]{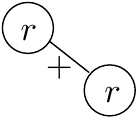}
	&  & $\left({0 \atop 0} {1 \atop 1}\right)$ or $\left({1 \atop 1} {0 \atop 0}\right)$ &
		$\left({0 \atop 0} {0 \atop 1}\right)$, $\left({0 \atop 1} {0 \atop 0}\right)$, $\left({1 \atop 1} {0 \atop 1}\right)$, or $\left({0 \atop 1} {1 \atop 1}\right)$  &
		$\left({0 \atop 0} {1 \atop 1}\right)$ or $\left({1 \atop 1} {0 \atop 0}\right)$ & 0R \\

      \midrule
    \multirow{2}{*}{\includegraphics[scale=0.7]{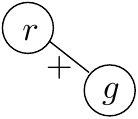}}
	& \cancel{B} & $\left({0 \atop 0} {1 \atop 1}\right)$ or $\left({1 \atop 1} {0 \atop 0}\right)$ &
		$\left({0 \atop 0} {0 \atop 1}\right)$, $\left({0 \atop 1} {0 \atop 0}\right)$, $\left({1 \atop 1} {0 \atop 1}\right)$, or $\left({0 \atop 1} {1 \atop 1}\right)$  &
		$\left({0 \atop 1} {0 \atop 1}\right)$ & {\bf MI } \\
	& \bf{R} & $\left({0 \atop 0} {1 \atop 1}\right)$ or $\left({1 \atop 1} {0 \atop 0}\right)$ &
		$\left({0 \atop 0} {0 \atop 1}\right)$, $\left({0 \atop 1} {0 \atop 0}\right)$, $\left({1 \atop 1} {0 \atop 1}\right)$, or $\left({0 \atop 1} {1 \atop 1}\right)$  &
		$\left({0 \atop 1} {1 \atop 0}\right)$ & {\bf 0R } \\

      \midrule
    \multirow{2}{*}{\includegraphics[scale=0.7]{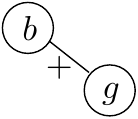}}
	& \bf{B} & $\left({0 \atop 0} {0 \atop 0}\right)$ or $\left({1 \atop 1} {1 \atop 1}\right)$ &
		$\left({0 \atop 0} {0 \atop 1}\right)$, $\left({0 \atop 1} {0 \atop 0}\right)$, $\left({1 \atop 1} {0 \atop 1}\right)$, or $\left({0 \atop 1} {1 \atop 1}\right)$  &
		$\left({0 \atop 1} {0 \atop 1}\right)$ & {\bf 0R } \\
	& \cancel{R} & $\left({0 \atop 0} {0 \atop 0}\right)$ or $\left({1 \atop 1} {1 \atop 1}\right)$ &
		$\left({0 \atop 0} {0 \atop 1}\right)$, $\left({0 \atop 1} {0 \atop 0}\right)$, $\left({1 \atop 1} {0 \atop 1}\right)$, or $\left({0 \atop 1} {1 \atop 1}\right)$  &
		$\left({0 \atop 1} {1 \atop 0}\right)$ & {\bf MI }\\

      \midrule
    \multirow{2}{*}{\includegraphics[scale=0.7]{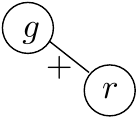}}
	& B & $\left({0 \atop 1} {0 \atop 1}\right)$ &
		$\left({0 \atop 0} {0 \atop 1}\right)$, $\left({0 \atop 1} {0 \atop 0}\right)$, $\left({1 \atop 1} {0 \atop 1}\right)$, or $\left({0 \atop 1} {1 \atop 1}\right)$  &
		$\left({0 \atop 0} {1 \atop 1}\right)$ or $\left({1 \atop 1} {0 \atop 0}\right)$ &  1R \\
	& R & $\left({0 \atop 1} {1 \atop 0}\right)$ &
		$\left({0 \atop 0} {0 \atop 1}\right)$, $\left({0 \atop 1} {0 \atop 0}\right)$, $\left({1 \atop 1} {0 \atop 1}\right)$, or $\left({0 \atop 1} {1 \atop 1}\right)$  &
		$\left({0 \atop 0} {1 \atop 1}\right)$ or $\left({1 \atop 1} {0 \atop 0}\right)$ &  0R \\

      \midrule
    \multirow{2}{*}{\includegraphics[scale=0.7]{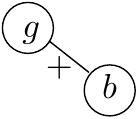}}
	& B & $\left({0 \atop 1} {0 \atop 1}\right)$ &
		$\left({0 \atop 0} {0 \atop 1}\right)$, $\left({0 \atop 1} {0 \atop 0}\right)$, $\left({1 \atop 1} {0 \atop 1}\right)$, or $\left({0 \atop 1} {1 \atop 1}\right)$  &
		$\left({0 \atop 0} {0 \atop 0}\right)$ or $\left({1 \atop 1} {1 \atop 1}\right)$ &  0R \\
	& R & $\left({0 \atop 1} {1 \atop 0}\right)$ &
		$\left({0 \atop 0} {0 \atop 1}\right)$, $\left({0 \atop 1} {0 \atop 0}\right)$, $\left({1 \atop 1} {0 \atop 1}\right)$, or $\left({0 \atop 1} {1 \atop 1}\right)$  &
		$\left({0 \atop 0} {0 \atop 0}\right)$ or $\left({1 \atop 1} {1 \atop 1}\right)$ &  1R \\

      \midrule
    \multirow{4}{*}{\includegraphics[scale=0.7]{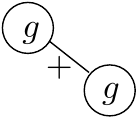}}
	& BB & $\left({0 \atop 1} {0 \atop 1}\right)$ &
		$\left({0 \atop 0} {0 \atop 1}\right)$, $\left({0 \atop 1} {0 \atop 0}\right)$, $\left({1 \atop 1} {0 \atop 1}\right)$, or $\left({0 \atop 1} {1 \atop 1}\right)$  &
		$\left({0 \atop 1} {0 \atop 1}\right)$  &  0R \\
	& BR & $\left({0 \atop 1} {0 \atop 1}\right)$ &
		$\left({0 \atop 0} {0 \atop 1}\right)$, $\left({0 \atop 1} {0 \atop 0}\right)$, $\left({1 \atop 1} {0 \atop 1}\right)$, or $\left({0 \atop 1} {1 \atop 1}\right)$  &
		$\left({0 \atop 1} {1 \atop 0}\right)$  &  1R \\
	& RB & $\left({0 \atop 1} {1 \atop 0}\right)$ &
		$\left({0 \atop 0} {0 \atop 1}\right)$, $\left({0 \atop 1} {0 \atop 0}\right)$, $\left({1 \atop 1} {0 \atop 1}\right)$, or $\left({0 \atop 1} {1 \atop 1}\right)$  &
		$\left({0 \atop 1} {0 \atop 1}\right)$  &  1R \\
	& RR & $\left({0 \atop 1} {1 \atop 0}\right)$ &
		$\left({0 \atop 0} {0 \atop 1}\right)$, $\left({0 \atop 1} {0 \atop 0}\right)$, $\left({1 \atop 1} {0 \atop 1}\right)$, or $\left({0 \atop 1} {1 \atop 1}\right)$  &
		$\left({0 \atop 1} {1 \atop 0}\right)$  &  0R \\

      \midrule
    \multirow{2}{*}{\includegraphics[scale=0.7]{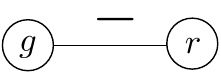}} & B & $\left({0 \atop 1} {0 \atop 1 }\right)$  &  $\left({0 \atop 0} {1 \atop 1}\right)$ or $\left({1 \atop 1} {0 \atop 0}\right)$ &  $\left({0 \atop 0} {0 \atop 1}\right)$, $\left({0 \atop 1} {0 \atop 0}\right)$, $\left({1 \atop 1} {0 \atop 1}\right)$, or $\left({0 \atop 1} {1 \atop 1}\right)$  &  0R \\
		 & R &  $\left({0 \atop 1} {1 \atop 0 }\right)$  &  $\left({0 \atop 0} {1 \atop 1}\right)$ or $\left({1 \atop 1} {0 \atop 0}\right)$ &  $\left({0 \atop 0} {0 \atop 1}\right)$, $\left({0 \atop 1} {0 \atop 0}\right)$, $\left({1 \atop 1} {0 \atop 1}\right)$, or $\left({0 \atop 1} {1 \atop 1}\right)$  &  1R \\

      \midrule
     \multirow{2}{*}{\includegraphics[scale=0.7]{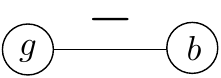}} & B & $\left({0 \atop 1} {0 \atop 1 }\right)$  &  $\left({0 \atop 0} {0 \atop 0}\right)$ or $\left({1 \atop 1} {1 \atop 1}\right)$ &  $\left({0 \atop 0} {0 \atop 1}\right)$, $\left({0 \atop 1} {0 \atop 0}\right)$, $\left({1 \atop 1} {0 \atop 1}\right)$, or $\left({0 \atop 1} {1 \atop 1}\right)$  &  1R \\
                            & R & $\left({0 \atop 1} {1 \atop 0 }\right)$  &  $\left({0 \atop 0} {0 \atop 0}\right)$ or $\left({1 \atop 1} {1 \atop 1}\right)$ &  $\left({0 \atop 0} {0 \atop 1}\right)$, $\left({0 \atop 1} {0 \atop 0}\right)$, $\left({1 \atop 1} {0 \atop 1}\right)$, or $\left({0 \atop 1} {1 \atop 1}\right)$  &  0R \\

    \multirow{3}{*}{\includegraphics[scale=0.7]{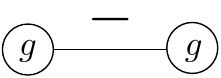}} & BB & $\left({0 \atop 1} {0 \atop 1 }\right)$  &  $\left({0 \atop 1} {0 \atop 1}\right)$ &  $\left({0 \atop 0} {0 \atop 1}\right)$, $\left({0 \atop 1} {0 \atop 0}\right)$, $\left({1 \atop 1} {0 \atop 1}\right)$, or $\left({0 \atop 1} {1 \atop 1}\right)$  &  1R \\
     & BR & $\left({0 \atop 1} {0 \atop 1 }\right)$  &  $\left({0 \atop 1} {1 \atop 0}\right)$ &  $\left({0 \atop 0} {0 \atop 1}\right)$, $\left({0 \atop 1} {0 \atop 0}\right)$, $\left({1 \atop 1} {0 \atop 1}\right)$, or $\left({0 \atop 1} {1 \atop 1}\right)$  &  0R \\
    				  & RR & $\left({0 \atop 1} {1 \atop 0 }\right)$  &  $\left({0 \atop 1} {1 \atop 0}\right)$  &  $\left({0 \atop 0} {0 \atop 1}\right)$, $\left({0 \atop 1} {0 \atop 0}\right)$, $\left({1 \atop 1} {0 \atop 1}\right)$, or $\left({0 \atop 1} {1 \atop 1}\right)$  &  1R \\

     \midrule
    \multirow{2}{*}{\includegraphics[scale=0.7]{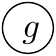}}
	& B & $\left({0 \atop 1} {0 \atop 1}\right)$ &
		$\left({0 \atop 1} {0 \atop 0}\right)$ or $\left({0 \atop 1} {1 \atop 1}\right)$ &
		$\left({0 \atop 1} {0 \atop 0}\right)$ or $\left({0 \atop 1} {1 \atop 1}\right)$ & 0R \\
	& R & $\left({0 \atop 1} {1 \atop 0}\right)$ &
		$\left({0 \atop 1} {0 \atop 0}\right)$ or $\left({0 \atop 1} {1 \atop 1}\right)$ &
		$\left({0 \atop 1} {0 \atop 0}\right)$ or $\left({0 \atop 1} {1 \atop 1}\right)$ & 0R \\

      \midrule
    \multirow{8}{*}{\includegraphics[scale=0.7]{case2l.pdf}}
	& \multirow{2}{*}{BBB}  & $\left({0 \atop 1} {0 \atop 0} {0 \atop 1}\right)$ &
		$\left({0 \atop 1} {0 \atop 1} {0 \atop 0}\right)$ &
		$\left({0 \atop 1} {0 \atop 1} {0 \atop 0}\right)$ & 0R \\
	   &  & $\left({0 \atop 1} {1 \atop 1} {0 \atop 1}\right)$ &
		$\left({0 \atop 1} {0 \atop 1} {1 \atop 1}\right)$ &
		$\left({0 \atop 1} {0 \atop 1} {1 \atop 1}\right)$ & 0R \\
	& \multirow{2}{*}{BBR}  & $\left({0 \atop 1} {0 \atop 0} {0 \atop 1}\right)$ &
		$\left({0 \atop 1} {0 \atop 1} {0 \atop 0}\right)$ &
		$\left({0 \atop 1} {1 \atop 0} {0 \atop 0}\right)$ & 2R \\
	   &  & $\left({0 \atop 1} {1 \atop 1} {0 \atop 1}\right)$ &
		$\left({0 \atop 1} {0 \atop 1} {1 \atop 1}\right)$ &
		$\left({0 \atop 1} {1 \atop 0} {1 \atop 1}\right)$ & 2R \\
	& \multirow{2}{*}{BRB}  & $\left({0 \atop 1} {0 \atop 0} {0 \atop 1}\right)$ &
		$\left({0 \atop 1} {1 \atop 0} {0 \atop 0}\right)$ &
		$\left({0 \atop 1} {0 \atop 1} {0 \atop 0}\right)$ & 1R \\
	   &  & $\left({0 \atop 1} {1 \atop 1} {0 \atop 1}\right)$ &
		$\left({0 \atop 1} {1 \atop 0} {1 \atop 1}\right)$ &
		$\left({0 \atop 1} {0 \atop 1} {1 \atop 1}\right)$ & 1R \\
	& \multirow{2}{*}{BRR}  & $\left({0 \atop 1} {0 \atop 0} {0 \atop 1}\right)$ &
		$\left({0 \atop 1} {1 \atop 0} {0 \atop 0}\right)$ &
		$\left({0 \atop 1} {1 \atop 0} {0 \atop 0}\right)$ & 1R \\
	   &  & $\left({0 \atop 1} {1 \atop 1} {0 \atop 1}\right)$ &
		$\left({0 \atop 1} {1 \atop 0} {1 \atop 1}\right)$ &
		$\left({0 \atop 1} {1 \atop 0} {1 \atop 1}\right)$ & 1R \\
	& \multirow{2}{*}{RBB}  & $\left({0 \atop 1} {0 \atop 0} {1 \atop 0}\right)$ &
		$\left({0 \atop 1} {0 \atop 1} {0 \atop 0}\right)$ &
		$\left({0 \atop 1} {0 \atop 1} {0 \atop 0}\right)$ & 1R \\
	   &  & $\left({0 \atop 1} {1 \atop 1} {1 \atop 0}\right)$ &
		$\left({0 \atop 1} {0 \atop 1} {1 \atop 1}\right)$ &
		$\left({0 \atop 1} {0 \atop 1} {1 \atop 1}\right)$ & 1R \\
	& \multirow{2}{*}{RBR}  & $\left({0 \atop 1} {0 \atop 0} {1 \atop 0}\right)$ &
		$\left({0 \atop 1} {0 \atop 1} {0 \atop 0}\right)$ &
		$\left({0 \atop 1} {1 \atop 0} {0 \atop 0}\right)$ & 1R \\
	   &  & $\left({0 \atop 1} {1 \atop 1} {1 \atop 0}\right)$ &
		$\left({0 \atop 1} {0 \atop 1} {1 \atop 1}\right)$ &
		$\left({0 \atop 1} {1 \atop 0} {1 \atop 1}\right)$ & 1R \\
	& \multirow{2}{*}{RRB}  & $\left({0 \atop 1} {0 \atop 0} {1 \atop 0}\right)$ &
		$\left({0 \atop 1} {1 \atop 0} {0 \atop 0}\right)$ &
		$\left({0 \atop 1} {0 \atop 1} {0 \atop 0}\right)$ & 2R \\
	   &  & $\left({0 \atop 1} {1 \atop 1} {1 \atop 0}\right)$ &
		$\left({0 \atop 1} {1 \atop 0} {1 \atop 1}\right)$ &
		$\left({0 \atop 1} {0 \atop 1} {1 \atop 1}\right)$ & 2R \\
	& \multirow{2}{*}{RRR}  & $\left({0 \atop 1} {0 \atop 0} {1 \atop 0}\right)$ &
		$\left({0 \atop 1} {1 \atop 0} {0 \atop 0}\right)$ &
		$\left({0 \atop 1} {1 \atop 0} {0 \atop 0}\right)$ & 0R \\
	   &  & $\left({0 \atop 1} {1 \atop 1} {1 \atop 0}\right)$ &
		$\left({0 \atop 1} {1 \atop 0} {1 \atop 1}\right)$ &
		$\left({0 \atop 1} {1 \atop 0} {1 \atop 1}\right)$ & 0R \\

      \midrule
    \multirow{8}{*}{\includegraphics[scale=0.7]{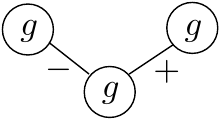}}
	& \multirow{2}{*}{BBB}  & $\left({0 \atop 1} {0 \atop 0} {0 \atop 1}\right)$ &
		$\left({0 \atop 1} {0 \atop 1} {1 \atop 1}\right)$ &
		$\left({0 \atop 1} {0 \atop 1} {1 \atop 1}\right)$ & 1R \\
	   &  & $\left({0 \atop 1} {1 \atop 1} {0 \atop 1}\right)$ &
		$\left({0 \atop 1} {0 \atop 1} {0 \atop 0}\right)$ &
		$\left({0 \atop 1} {0 \atop 1} {0 \atop 0}\right)$ & 1R \\
	& \multirow{2}{*}{BBR}  & $\left({0 \atop 1} {0 \atop 0} {0 \atop 1}\right)$ &
		$\left({0 \atop 1} {0 \atop 1} {1 \atop 1}\right)$ &
		$\left({0 \atop 1} {1 \atop 0} {1 \atop 1}\right)$ & 1R \\
	   &  & $\left({0 \atop 1} {1 \atop 1} {0 \atop 1}\right)$ &
		$\left({0 \atop 1} {0 \atop 1} {0 \atop 0}\right)$ &
		$\left({0 \atop 1} {1 \atop 0} {0 \atop 0}\right)$ & 1R \\
	& \multirow{2}{*}{BRB}  & $\left({0 \atop 1} {0 \atop 0} {0 \atop 1}\right)$ &
		$\left({0 \atop 1} {1 \atop 0} {1 \atop 1}\right)$ &
		$\left({0 \atop 1} {0 \atop 1} {1 \atop 1}\right)$ & 2R \\
	   &  & $\left({0 \atop 1} {1 \atop 1} {0 \atop 1}\right)$ &
		$\left({0 \atop 1} {1 \atop 0} {0 \atop 0}\right)$ &
		$\left({0 \atop 1} {0 \atop 1} {0 \atop 0}\right)$ & 2R \\
	& \multirow{2}{*}{BRR}  & $\left({0 \atop 1} {0 \atop 0} {0 \atop 1}\right)$ &
		$\left({0 \atop 1} {1 \atop 0} {1 \atop 1}\right)$ &
		$\left({0 \atop 1} {1 \atop 0} {1 \atop 1}\right)$ & 0R \\
	   &  & $\left({0 \atop 1} {1 \atop 1} {0 \atop 1}\right)$ &
		$\left({0 \atop 1} {1 \atop 0} {0 \atop 0}\right)$ &
		$\left({0 \atop 1} {1 \atop 0} {0 \atop 0}\right)$ & 0R \\
	& \multirow{2}{*}{RBB}  & $\left({0 \atop 1} {0 \atop 0} {1 \atop 0}\right)$ &
		$\left({0 \atop 1} {0 \atop 1} {1 \atop 1}\right)$ &
		$\left({0 \atop 1} {0 \atop 1} {1 \atop 1}\right)$ & 0R \\
	   &  & $\left({0 \atop 1} {1 \atop 1} {1 \atop 0}\right)$ &
		$\left({0 \atop 1} {0 \atop 1} {0 \atop 0}\right)$ &
		$\left({0 \atop 1} {0 \atop 1} {0 \atop 0}\right)$ & 0R \\
	& \multirow{2}{*}{RBR}  & $\left({0 \atop 1} {0 \atop 0} {1 \atop 0}\right)$ &
		$\left({0 \atop 1} {0 \atop 1} {1 \atop 1}\right)$ &
		$\left({0 \atop 1} {1 \atop 0} {1 \atop 1}\right)$ & 2R \\
	   &  & $\left({0 \atop 1} {1 \atop 1} {1 \atop 0}\right)$ &
		$\left({0 \atop 1} {0 \atop 1} {0 \atop 0}\right)$ &
		$\left({0 \atop 1} {1 \atop 0} {0 \atop 0}\right)$ & 2R \\
	& \multirow{2}{*}{RRB}  & $\left({0 \atop 1} {0 \atop 0} {1 \atop 0}\right)$ &
		$\left({0 \atop 1} {1 \atop 0} {1 \atop 1}\right)$ &
		$\left({0 \atop 1} {0 \atop 1} {1 \atop 1}\right)$ & 1R \\
	   &  & $\left({0 \atop 1} {1 \atop 1} {1 \atop 0}\right)$ &
		$\left({0 \atop 1} {1 \atop 0} {0 \atop 0}\right)$ &
		$\left({0 \atop 1} {0 \atop 1} {0 \atop 0}\right)$ & 1R \\
	& \multirow{2}{*}{RRR}  & $\left({0 \atop 1} {0 \atop 0} {1 \atop 0}\right)$ &
		$\left({0 \atop 1} {1 \atop 0} {1 \atop 1}\right)$ &
		$\left({0 \atop 1} {1 \atop 0} {1 \atop 1}\right)$ & 1R \\
	   &  & $\left({0 \atop 1} {1 \atop 1} {1 \atop 0}\right)$ &
		$\left({0 \atop 1} {1 \atop 0} {0 \atop 0}\right)$ &
		$\left({0 \atop 1} {1 \atop 0} {0 \atop 0}\right)$ & 1R \\

      \midrule
    \multirow{4}{*}{\includegraphics[scale=0.7]{case1c.pdf}}
	& \multirow{2}{*}{BB} & $\left({0 \atop 1} {0 \atop 0} {0 \atop 1}\right)$ &
		$\left({0 \atop 1} {0 \atop 1} {0 \atop 0}\right)$ &
		$\left({0 \atop 1} {0 \atop 0} {0 \atop 0}\right)$ &  1R \\
	& & $\left({0 \atop 1} {1 \atop 1} {0 \atop 1}\right)$ &
		$\left({0 \atop 1} {0 \atop 1} {1 \atop 1}\right)$ &
		$\left({0 \atop 1} {1 \atop 1} {1 \atop 1}\right)$  &  1R \\

	& \multirow{2}{*}{BR} & $\left({0 \atop 1} {0 \atop 0} {0 \atop 1}\right)$ &
		$\left({0 \atop 1} {1 \atop 0} {0 \atop 0}\right)$ &
		$\left({0 \atop 1} {0 \atop 0} {0 \atop 0}\right)$ &  0R \\
	& & $\left({0 \atop 1} {1 \atop 1} {0 \atop 1}\right)$ &
		$\left({0 \atop 1} {1 \atop 0} {1 \atop 1}\right)$ &
		$\left({0 \atop 1} {1 \atop 1} {1 \atop 1}\right)$  &  0R \\

	& \multirow{2}{*}{RB} & $\left({0 \atop 1} {0 \atop 0} {1 \atop 0}\right)$ &
		$\left({0 \atop 1} {0 \atop 1} {0 \atop 0}\right)$ &
		$\left({0 \atop 1} {0 \atop 0} {0 \atop 0}\right)$ &  0R \\
	& & $\left({0 \atop 1} {1 \atop 1} {1 \atop 0}\right)$ &
		$\left({0 \atop 1} {0 \atop 1} {1 \atop 1}\right)$ &
		$\left({0 \atop 1} {1 \atop 1} {1 \atop 1}\right)$  &  0R \\

	& \multirow{2}{*}{RR} & $\left({0 \atop 1} {0 \atop 0} {1 \atop 0}\right)$ &
		$\left({0 \atop 1} {1 \atop 0} {0 \atop 0}\right)$ &
		$\left({0 \atop 1} {0 \atop 0} {0 \atop 0}\right)$ &  1R \\
	& & $\left({0 \atop 1} {1 \atop 1} {1 \atop 0}\right)$ &
		$\left({0 \atop 1} {1 \atop 0} {1 \atop 1}\right)$ &
		$\left({0 \atop 1} {1 \atop 1} {1 \atop 1}\right)$  &  1R \\

	\midrule
    \multirow{4}{*}{\includegraphics[scale=0.7]{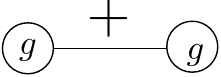}}
	& \multirow{2}{*}{BB} & $\left({0 \atop 1} {0 \atop 0} {0 \atop 1}\right)$ &
		$\left({0 \atop 1} {0 \atop 1} {1 \atop 1}\right)$ &
		$\left({0 \atop 1} {0 \atop 0} {1 \atop 1}\right)$ &  0R \\
	& & $\left({0 \atop 1} {1 \atop 1} {0 \atop 1}\right)$ &
		$\left({0 \atop 1} {0 \atop 1} {0 \atop 0}\right)$ &
		$\left({0 \atop 1} {1 \atop 1} {0 \atop 0}\right)$  &  0R \\

	& \multirow{2}{*}{BR} & $\left({0 \atop 1} {0 \atop 0} {0 \atop 1}\right)$ &
		$\left({0 \atop 1} {1 \atop 0} {1 \atop 1}\right)$ &
		$\left({0 \atop 1} {0 \atop 0} {1 \atop 1}\right)$ &  1R \\
	& & $\left({0 \atop 1} {1 \atop 1} {0 \atop 1}\right)$ &
		$\left({0 \atop 1} {1 \atop 0} {0 \atop 0}\right)$ &
		$\left({0 \atop 1} {1 \atop 1} {0 \atop 0}\right)$  &  1R \\

	& \multirow{2}{*}{RB} & $\left({0 \atop 1} {0 \atop 0} {1 \atop 0}\right)$ &
		$\left({0 \atop 1} {0 \atop 1} {1 \atop 1}\right)$ &
		$\left({0 \atop 1} {0 \atop 0} {1 \atop 1}\right)$ &  1R \\
	& & $\left({0 \atop 1} {1 \atop 1} {1 \atop 0}\right)$ &
		$\left({0 \atop 1} {0 \atop 1} {0 \atop 0}\right)$ &
		$\left({0 \atop 1} {1 \atop 1} {0 \atop 0}\right)$  &  1R \\

	& \multirow{2}{*}{RR} & $\left({0 \atop 1} {0 \atop 0} {1 \atop 0}\right)$ &
		$\left({0 \atop 1} {1 \atop 0} {1 \atop 1}\right)$ &
		$\left({0 \atop 1} {0 \atop 0} {1 \atop 1}\right)$ &  0R \\
	& & $\left({0 \atop 1} {1 \atop 1} {1 \atop 0}\right)$ &
		$\left({0 \atop 1} {1 \atop 0} {0 \atop 0}\right)$ &
		$\left({0 \atop 1} {1 \atop 1} {0 \atop 0}\right)$  &  0R \\

      \midrule
    \multirow{2}{*}{\includegraphics[scale=0.7]{case1b.pdf}}
	& \multirow{2}{*}{B}  & $\left({0 \atop 1} {0 \atop 1}\right)$ &
		$\left({0 \atop 1} {0 \atop 0}\right)$ &
		$\left({0 \atop 0} {0 \atop 1}\right)$ & 1R \\
	   &  & $\left({0 \atop 1} {0 \atop 1}\right)$ &
		$\left({0 \atop 1} {1 \atop 1}\right)$ &
		$\left({1 \atop 1} {0 \atop 1}\right)$ & 1R \\
	& \multirow{2}{*}{R}  & $\left({0 \atop 1} {1 \atop 0}\right)$ &
		$\left({0 \atop 1} {0 \atop 0}\right)$ &
		$\left({0 \atop 0} {0 \atop 1}\right)$ & 0R \\
	   &  & $\left({0 \atop 1} {1 \atop 0}\right)$ &
		$\left({0 \atop 1} {1 \atop 1}\right)$ &
		$\left({1 \atop 1} {0 \atop 1}\right)$ & 0R \\

      \midrule
    \multirow{2}{*}{\includegraphics[scale=0.7]{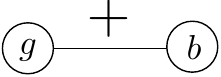}}
	& \multirow{2}{*}{B}  & $\left({0 \atop 1} {0 \atop 1}\right)$ &
		$\left({0 \atop 1} {0 \atop 0}\right)$ &
		$\left({1 \atop 1} {0 \atop 1}\right)$ & 0R \\
	   &  & $\left({0 \atop 1} {0 \atop 1}\right)$ &
		$\left({0 \atop 1} {1 \atop 1}\right)$ &
		$\left({0 \atop 0} {0 \atop 1}\right)$ & 0R \\
	& \multirow{2}{*}{R}  & $\left({0 \atop 1} {1 \atop 0}\right)$ &
		$\left({0 \atop 1} {0 \atop 0}\right)$ &
		$\left({1 \atop 1} {0 \atop 1}\right)$ & 1R \\
	   &  & $\left({0 \atop 1} {1 \atop 0}\right)$ &
		$\left({0 \atop 1} {1 \atop 1}\right)$ &
		$\left({0 \atop 0} {0 \atop 1}\right)$ & 1R \\

\newpage
   \multirow{2}{*}{\includegraphics[scale=0.7]{case4.pdf} child}
	& \multirow{2}{*}{B} & $\left({0 \atop 0} {0 \atop 1}\right)$ &
		$\left({1 \atop 1} {0 \atop 1}\right)$  &
		$\left({0 \atop 1} {0 \atop 1}\right)$ & 0R \\
	&  & $\left({0 \atop 0} {0 \atop 1}\right)$ &
		$\left({0 \atop 1} {1 \atop 1}\right)$  &
		$\left({0 \atop 1} {0 \atop 1}\right)$ & 0R \\

	& \multirow{2}{*}{R} & $\left({0 \atop 0} {0 \atop 1}\right)$ &
		$\left({1 \atop 1} {0 \atop 1}\right)$  &
		$\left({0 \atop 1} {1 \atop 0}\right)$ & 0R \\
	&  & $\left({0 \atop 0} {0 \atop 1}\right)$ &
		$\left({0 \atop 1} {1 \atop 1}\right)$  &
		$\left({0 \atop 1} {1 \atop 0}\right)$ & 0R \\

	& \multirow{2}{*}{B} & $\left({0 \atop 1} {0 \atop 0}\right)$ &
		$\left({1 \atop 1} {0 \atop 1}\right)$  &
		$\left({0 \atop 1} {0 \atop 1}\right)$ & 0R \\
	&  & $\left({0 \atop 1} {0 \atop 0}\right)$ &
		$\left({0 \atop 1} {1 \atop 1}\right)$  &
		$\left({0 \atop 1} {0 \atop 1}\right)$ & 0R \\

	& \multirow{2}{*}{R}  & $\left({0 \atop 1} {0 \atop 0}\right)$ &
		$\left({1 \atop 1} {0 \atop 1}\right)$  &
		$\left({0 \atop 1} {1 \atop 0}\right)$ & 0R \\
	&  & $\left({0 \atop 1} {0 \atop 0}\right)$ &
		$\left({0 \atop 1} {1 \atop 1}\right)$  &
		$\left({0 \atop 1} {1 \atop 0}\right)$ & 0R \\

	& \multirow{2}{*}{\cancel{B}}  & $\left({0 \atop 0} {0 \atop 1}\right)$ &
		$\left({0 \atop 1} {0 \atop 0}\right)$  &
		$\left({0 \atop 1} {0 \atop 1}\right)$ & {\bf MI } \\
	&  & $\left({1 \atop 1} {0 \atop 1}\right)$ &
		$\left({0 \atop 1} {1 \atop 1}\right)$  &
		$\left({0 \atop 1} {0 \atop 1}\right)$ & {\bf MI } \\

	& \multirow{2}{*}{\bf{R}} & $\left({0 \atop 0} {0 \atop 1}\right)$ &
		$\left({0 \atop 1} {0 \atop 0}\right)$  &
		$\left({0 \atop 1} {1 \atop 0}\right)$ & {\bf 0R } \\
	&  & $\left({1 \atop 1} {0 \atop 1}\right)$ &
		$\left({0 \atop 1} {1 \atop 1}\right)$  &
		$\left({0 \atop 1} {1 \atop 0}\right)$ & {\bf 0R } \\
  \label{tab:mrgraphproperty1}
\end{longtable}

\end{center}

\end{document}